\documentclass[twocolumn,showpacs,preprintnumbers,superscriptaddress,amsmath,amssymb,floatfix,prb]{revtex4}

\usepackage{graphicx}
\usepackage{dcolumn}
\usepackage{psfrag}
\usepackage{epsfig}
\usepackage{amsmath}
\usepackage{amssymb}
\usepackage{color}
\usepackage{mathbbol}
\usepackage{subfigure}
\usepackage{epstopdf,txfonts}
\usepackage{hyperref} 
\usepackage{epstopdf}
\usepackage[multidot]{grffile}

\newcommand{\kk}{\mathbf{k}}
\newcommand{\rr}{\mathbf{r}}
\newcommand{\trace}{{\rm Tr}}

\begin{document}

\title{The quantum phase transition and correlations in the multi-spin-boson model}

\author{Andr\'{e} Winter}
\affiliation{Theoretische Physik, Universit\"{a}t des Saarlandes, D-66123 Saarbr\"{u}cken, Germany}

\author{Heiko Rieger}
\affiliation{Theoretische Physik, Universit\"{a}t des Saarlandes, D-66123 Saarbr\"{u}cken, Germany}

\date{\today}
\begin{abstract}
We consider multiple non-interacting quantum mechanical two-level 
systems coupled to a common bosonic bath and study its quantum 
phase transition with Monte Carlo simulations using a continuous 
imaginary time cluster algorithm. The common bath induces an effective
ferromagnetic interaction between the otherwise independent two-level
systems, which can be quantified by an effective interaction strength.
For degenerate energy levels above a critical value of the bath 
coupling strength $\alpha$ all two-level systems freeze into the 
same state and the critical value $\alpha_c$ decreases asymptotically
as $1/N$ with increasing $N$.
For a finite number, $N$, of two-level systems the quantum phase 
transition (at zero temperature) is in the same universality class 
as the single spin-boson model, in the limit $N\to\infty$ the
system shows mean-field critical behavior independent of
the power of the spectral function of the bosonic bath.
We also study the influence of a spatial separation of the spins in a bath 
of bosonic modes with linear dispersion relation on the location and
characteristics of the phase transition as well as on correlations between 
the two-level systems.
\end{abstract}

\maketitle

\section{Introduction}
\label{sec:intro}
The single spin-boson model \cite{CLM,Weissbook} describes a two-level system, like a spin-1/2 or a q-bit, which is linearly coupled to a bath 
of bosonic modes. 
In spite of its simplicity the system shows a multifaceted behavior
in statics, dynamics and quantum criticality, for which reason it
became a paradigmatic model in the field of quantum dissipation.
For Ohmic and sub-Ohmic dissipation and degenerate energy of the
two states the two-level system shows a transition at zero temperature 
from a delocalized state (tunneling between the two states) at weak 
coupling to the bath to a localized, classical behavior (frozen 
in one state). The characteristics, in particular the universality 
class, of this quantum phase transition is identical to the one 
of the thermal transition in the classical Ising chain with long-range interactions\cite{Winter2009,Alvermann2009,Zhang2010}.

An interesting question is: what happens to this transition if
one couples several two-level systems, like impurities or q-bits, 
to a bosonic dissipative bath? The simplest generalization of this kind 
is a pair of two-level systems, which has been studied the first time
in [\onlinecite{Stamp1998}] and subsequently by
[\onlinecite{Governale2001,Thorwart2002,Garst2004,Nagele2008,Nagele2010,
McCutcheon2010,LeHur2010,Bonart2013}].
In many studies of the two-spin-boson model an additional ferromagnetic
coupling between the two-level systems of spins has been taken into account.
Arranging $N$ two-level systems in a chain with uniform ferromagnetic 
nearest neighbor interactions and coupling each of it to its own
bosonic bath yields the transverse Ising chain with dissipation
that has been studied for an Ohmic bath in [\onlinecite{Werner2005}]. 
There it was shown that in the limit $N\to\infty$ a new quantum phase transition triggered by the bath coupling
strength emerges, which is different from the quantum phase transition
in the transverse Ising chain without dissipation and different from
the single spin-boson model. If the ferromagnetic couplings
between the two-level systems are not uniform but random
the random transverse Ising chain with dissipation is obtained,
which has been studied in [\onlinecite{Cugliandolo2005,Schehr2006}].
Here not a sharp phase transition occurs but a smeared transition
in which connected clusters of two-level systems undergo 
separate quantum phase transitions at different coupling 
strengths \cite{Schehr2006,Vojta2008}.

These earlier investigations for arbitrary two-level system number 
$N$ assume an independent bath for each two-level system. This might
not always represent correctly the physical situation, in which a 
single bath for all two-level systems might be more appropriate -- 
as for instance in cold atom or trapped ion setups. In addition 
a direct coupling between the two-level systems might be absent.
This is the situation in which we address in this paper by 
studying the multi-spin-boson model (MSBM) with a single bosonic bath.
One expects that in spite of the absence of a direct interaction 
between the two-level systems the common bath will mediate an
effective interaction between the two-level systems that enhances
the tendency towards localization and thus decreases the critical
bath coupling strength. 

For $N=2$ and Ohmic bath a recent variational calculation \cite{McCutcheon2010} 
obtained a critical coupling of $\alpha_c\approx1/2$, whereas 
a numerical renormalization group (NRG) calculation \cite{LeHur2010}
predicted (for vanishing direct coupling between the two-level 
systems) $\alpha_c\approx1/4$. Regarding such a large deviation
between the naive mean-field prediction of $\alpha_c(N)\propto 1/N$
and the NRG result for $N=2$ a close look at larger values for $N$
using exact methods like quantum Monte Carlo seems worthwhile. 
Therefore in this paper we analyze quantitatively in which way 
the location and characteristics of the quantum phase 
transition varies with the number $N$ and the separation of 
two-level systems and also shed light on the correlations
between the two-level systems mediated by the bath in the different phases.

The paper is organized as follows: In section II we introduce the
MSBM including its path integral representation and discuss the
zero-tunneling limit as well as the mean-field limit. The quantum
phase transition in the MSBM model is analyzed in section III for general
number of two-level systems, $N$, first theoretically and then with
the help of extensive quantum Monte Carlo simulations. In Section IV
the spin-spin correlations in the MSBM in the different
phases are discussed and in section V the effect of spatial 
separations of the two-level systems in the MSBM is analyzed.
Section VI contains a summary and a discussion of open questions.
Several details of the calculations as well as a detailed
description of the quantum Monte Carlo algorithm are
deferred to three appendices.

\section{The model and mean field theory}
\label{sec:nimp}

For completeness and for further reference we first summarize 
the characteristics of the quantum phase transition in the 
single spin-boson model. Its Hamiltonian is given by (\(\hbar = 1\))
\begin{align}
\label{Ham_SBM}
H = \frac{\Delta}{2}{\sigma^x} -  \frac{\epsilon}{2}{\sigma^z} + \sum_{k} {\omega_k}\, {a_k^\dagger} {a_k} + H_{I}
\end{align}
where \(\sigma^{x,z}\) are the Pauli spin-1/2 operators, \({a_k^\dagger}\)(\(a_k\)) are the bosonic creation (annihilation) operators, \(\Delta\) the bare tunneling amplitude, \(\omega_k\) is the frequency of the \({k-}\)th bath mode, $\epsilon$ a energy-bias of one of the two states, which we set
to zero ($\epsilon=0$). The interaction term
\begin{align}
\label{Ham_SBM_HI}
H_{I} =\frac{1}{2}\sum_{k}  \left(\lambda_k a^{\dagger}_k + \lambda^{*}_{k} a_{k} \right) \sigma^z
\end{align}
describes the linear coupling between the spin and its environment. The bath spectrum \(J(\omega)=\pi\sum_k |\lambda_k|^2  \delta(\omega-\omega_k)\) has 
a power-law form
\(J(\omega) = 2 \pi \alpha\omega^{s}\omega_c^{1-s} \theta(\omega_c-\omega)\), with the bath exponent \(s>0\), the coupling strength \(\alpha\) and a sharp cutoff at the frequency \(\omega_c\).

The most prominent case is the Ohmic spectrum, \(s=1\), which has a phase transition in the Kosterlitz-Thouless universality class between a delocalized phase with finite effective renormalized tunneling amplitude \(\Delta_{\mathrm{eff}}\) and a localized phase, where tunneling is completely suppressed, at \(\alpha_c = 1 + \mathcal{O}\left(\frac{\Delta}{\omega_c}\right)\) \cite{CLM,Weissbook}. 
In the sub-Ohmic regime (\(0 < s < 1\)), the phase transition is of 
second order, which is described by a Gaussian fixed point with "classical" 
exponents for 
\(0 < s \leq \frac{1}{2}\) and has non-classical, \(s\)-dependent 
critical exponents for \(\frac{1}{2} < s <1 \)\cite{Winter2009,Alvermann2009}.

In this paper we consider the generalization of (\ref{Ham_SBM}) to
$N$ two-level systems coupled to a common bath. The Hamiltonian of
this multi-spin-boson model (MSBM) is given by
\begin{align}
\label{Ham_MSBM}
H = \sum_{m=1}^{N} \left( \frac{\Delta}{2}{\sigma_{m}^x} -  \frac{\epsilon}{2}{\sigma_{m}^z}\right) + \; \sum_{k} {\omega_k}\, {a_k^\dagger} {a_k} + H_{I}
\end{align}
with the interaction term
\begin{align}
\label{HI_MSBM}
H_{I} =\sum_{m=1}^{N} \frac{1}{2}\sum_{k=-\infty}^{\infty}  \left(\lambda_k a^{\dagger}_k  + \lambda^{*}_{k} a_{k}\right) \sigma_m^z
\end{align}
Note that this setup is different from the dissipative spin-chain \cite{Werner2005,Schehr2006,Vojta2008}, where every spin is embedded in its own dissipative bath. The main effect of the non-Markovian common bath is that the spins are sharing the same polarization induced to the bosonic degrees of freedom. From this picture one gets the intuition that the localization of the multi-spin system is enhanced in comparison to the single spin case. The delocalized-to-localized transition point \(\alpha_c\) is hence expected to become lowered as the number of contributing spins is increased.

\subsection{Path integral representation}
\label{sec:pathintegral}
The partition function of the single-spin-boson model can be 
exactly written as a path integral\cite{Winter2009}, which 
is straightforward to generalize to the multi-spin-boson model,
see Appendix \ref{sec:deriv}. The partition function $Z$ of the quantum 
system is expressed as a sum over all possible spin-\(\frac{1}{2}\) worldlines \(s(\tau) \in \{-1,1\}\) with \(0 \leq \tau < \beta\)
\begin{align}
\label{eq:partition_func}
Z = \trace \left( \exp (- \beta H)\right) \propto \int \prod_{m=1}^{N}\mathcal{D}\left[s_m(\tau)\right]\; \exp(-S)\;.
\end{align}
The position of the kinks, where \(s(\tau)\) changes its sign, are 
Poissonian distributed and the effective action (for symmetric states, 
i.e.\ $\epsilon=0$) is
\begin{align}
\label{eq:action}
S =& -\frac{1}{2 \beta} \sum_{m=1}^N \sum_{m'=1}^N \Bigg[\int_0^{\beta} \int_0^{\tau}  s_m(\tau) s_{m'}(\tau') K_{\beta}(\tau - \tau')  \mathrm{d}\tau' \mathrm{d}\tau \Bigg]\;,
\end{align}
where the integral kernel is given by
\begin{align}
\label{eq:kernel}
K_{\beta}(\tau) &=\int_0^{\omega_c} \frac{J(\omega)}{\pi \omega} \left\{\frac{\beta \omega}{2} \frac{\cosh\left(\frac{\beta \omega}{2}\left(1-  \frac{2 \tau }{\beta} \right) \right)}{\sinh\left( \frac{\beta \omega}{2} \right)} \right\} \mathrm{d}\omega\;.
\end{align}
This function decays algebraically as \(\propto \tau^{- 1 - s}\) for \(\beta \rightarrow \infty\) and the finite cutoff frequency \(\omega_c\) ensures the convergence at \({\tau = 0}\).

The path integral representations of 
thermodynamic observables like the order parameter
(magnetization) $m$, the susceptibility $\chi$, and 
the dimensionless moment ratio $Q$ of the MSBM are then 
given by
\begin{eqnarray}
m&=\langle \sigma_i^z\rangle_{\rm MSBM}&
=\langle\vert M_i\vert\rangle_{\rm PI}\\
\chi&=2 \frac{\partial m}{\partial \epsilon}\vert_{\epsilon=0}&=
\beta \left(\langle M_i^2\rangle_{\rm PI}-m^2\right)\\
Q&=&\; \langle M_i^2\rangle_{\rm PI}^2/\langle M_i^4\rangle_{\rm PI}
\end{eqnarray}
where $M_i$ is the magnetization of the $i-$th worldline
\begin{equation}
M_i[s_i(\tau)]=\beta^{-1}\int_0^\beta d\tau \,s_i(\tau)
\end{equation}
(note that the observables $m$, $\chi$, and $Q$ are independent of the
index $i=1,\ldots,N$) and $\langle\cdots\rangle$ denotes 
expectation values with respect to the classical action
\begin{equation}
\langle{\cal O}\rangle_{\rm PI} = 
\int \prod_{m=1}^{N}\mathcal{D}\left[s_m(\tau)\right]\;{\cal O}\;\exp(-S)\;.
\end{equation}
Our quantum Monte Carlo cluster algorithm\cite{Winter2009} (see also Appendix \ref{sec:algorithm}) simply samples stochastically worldline configurations according to the
probability measure and the classical action.

\subsection{Zero tunneling limit}
\label{sec:zerotunnelinglimit}

Let us consider the zero tunneling limit \(\Delta = 0\). This simplifies the terms in (\ref{eq:action}) and (\ref{eq:kernel}) drastically, because no kinks can occur and hence the spin-worldlines are only spin variables \(s_m \in \{-1,1\}\) without any \(\tau\)-dependence. The path integral becomes to a sum over these variables and the double integral over the curly brackets in Eq. (\ref{eq:kernel}) is just \(\beta^2/2\). The remaining frequency integral over the spectral function can performed elementary and the resulting partition function becomes to
\begin{align}
\label{eq:action_D0}
Z = \sum_{s_1, \dots, s_N=\pm1} \exp \left( \frac{\beta}{2} \sum_{m=1}^N \sum_{m'=1}^N s_m s_{m'} \frac{\alpha \omega_c}{s} \right)\;.
\end{align}
This partition function is identical to the one for
a classical Ising model in which $N$ Ising spins 
interact with each other with the ferromagnetic coupling strength
\begin{align}
\label{eq:Ising_coupling}
J = \omega_c \alpha / s
\end{align}
between all distinct spin pairs. The thermodynamic limit (\(N \rightarrow \infty\)) is not yet well-defined, because the energy would grow quadratic with the system size. To ensure a proper limit, we introduce the scaled coupling strength
\begin{align}
  \label{eq:alpha_tilde}
  \tilde{\alpha} = \alpha N\;.
\end{align}
Thus, for zero tunneling strength the MSBM model is identical to the
exactly solvable classical mean-field model \cite{Baxterbook}, which provides a paramagnetic to ferromagnetic phase transition of mean-field type. Adapted to the present nomenclature in (\ref{eq:action_D0}), the transition point is
\begin{align}
  \label{eq:alpha_D0}
\tilde{\alpha_c}   = \frac{s}{\omega_c} T\;.
\end{align}


\subsection{Mean-field theory}
\label{sec:finiteT}
The mean-field approximation consists in neglecting terms that
are quadratic in the fluctuations of the order parameter
\(m = \langle \sigma^z \rangle\), which means one sets 
\((\sigma^z_i - m)(\sigma^z_j - m) \approx 0\). 
Since the integral in Eq. (\ref{eq:kernel}) can be computed
for $\tau=\beta$ and yields \(K^{(1)}(\beta)=2 \alpha \omega_c / s\), 
the action (\ref{eq:action}) is in mean-field approximation 
\begin{align}
\label{eq:action_MFT}
& S_{MF} = \sum_{n=1}^N \left(\frac{\beta \tilde{\alpha} \omega_c }{2s} m^2 - \frac{1}{2} \int_0^{\beta} \left(\epsilon + \frac{2 \tilde{\alpha} \omega_c}{s} m \right)s_{n}(\tau) \right)
\end{align}
Solving the self-consistent equation \(\frac{m}{2}=\frac{1}{N \beta} \frac{\partial \ln(Z)}{\partial \epsilon}\), one obtains a critical coupling strength
\begin{equation}
\label{eq:alpha_c_MFT}
\tilde{\alpha}_c = \frac{s \Delta}{2 \tanh(\beta \Delta / 2)\omega_c}\;,
\end{equation}
which is identical to the result in Ref. [\onlinecite{HouTong2010}] for 
\(N=1\). Note that the mean-field approximation predicts a phase transition
also at non-vanishing temperatures ($T>0$, i.e.\ $\beta<\infty$), which is 
absent in the MSBM for any finite $N$ (see next subsection).
Within the mean-filed approximation the dependence of the critical coupling 
$\alpha_c$ on the number of spins $N$ is fully compensated by the rescaling of the coupling strength (\ref{eq:alpha_tilde}) and Eq. (\ref{eq:alpha_c_MFT}) predicts a phase transition at the 
temperature
\begin{align}
\label{eq:T_alpha_c_MFT}
T_c = \frac{\Delta}{2 \operatorname{artanh}\left(s \Delta /(2 \tilde{\alpha}_c \omega_c)  \right)}\;.
\end{align}
At zero temperature one gets \(\tilde{\alpha}_c(T \rightarrow 0) = s \Delta /(2 \omega_c)\).
For zero tunneling one obtains $T_c(\Delta\to0)=\tilde{\alpha}\omega_c/s$,
which is identical to the exact result (\ref{eq:alpha_D0}) as expected, since
for zero tunneling the spins do not fluctuate and therefore the mean-field approximation is exact. 

Also in the limit $N\to\infty$ one expects mean-field theory to be
exact (see Appendix~\ref{sec:mean-field_solution}), which we checked
with extensive QMC simulations by finite $N$ scaling at fixed temperatures
(see next subsection).

\section{The quantum phase transition in the MSBM}
\label{sec:QPT_MSBM}

In this section, we study the quantum phase transition
(at $T=0$, i.e.\ $\beta=\infty$, and vanishing bias $\epsilon=0$) 
of the MSBM for different values of $N$. First we argue in the 
next subsection that
the universality class is independent of $N$, then we 
validate this prediction with 
finite $\beta$ scaling of QMC data.

\subsection{Universality class: Theoretical considerations}

\label{sec:QPT_MSBM_theory}
The universality class of the transition of the MSBM for general $N$ is not known and before we embark on a QMC study we argue in the following that for any finite $N$ the transition is in the same universality class as 
the single-spin-boson model ($N=1$). Consider a single worldline of a \(1/2\)-spin in a transverse field \(\Delta \sigma^x /2\). The field defines a characteristic length \(2/\Delta\) between two kinks \cite{fn:seg_length}. The mean number of segments of the worldline is hence \(L=\beta \Delta/2\). Therefore, the partition function (\ref{eq:partition_func}), (\ref{eq:action}) can approximately be written for large  \(\beta\) in a discretized form
\begin{align}
\label{eq:partition_discret}
Z = \sum_{\{s_{m,i}\}} \exp \left[ \frac{1}{2 \beta} \sum_{m,m'}^N \frac{1}{2}\sum_{i,j}^{L} s_{m,i} s_{m',j} K_{\beta}\left(|j-i|\frac{2}{\Delta}\right) \left(\frac{2}{\Delta}\right)^2 \right]\;,
\end{align}
where \(s_{m,i} \in \{-1,1\}\) denotes the spin variables of the \(m-\)th worldline at the \(i-\)th segment. Basically, the two integrals in (\ref{eq:action}) are replaced by sums. The spin variables can be summed up to a large spin variable \(M_i = \sum_m s_{m,i} \in \{-N,-N+2, \dots N\}\) involving a combinatorial factor
\begin{align}
Z = \sum_{\{M_{i}\}} \prod_{i=1}^{L} {N \choose \frac{N+M_i}{2}}\exp \left[ \frac{1}{4 \beta} \sum_{i,j}^{L} M_{i} M_{j} K_{\beta}\left(|j-i|\frac{2}{\Delta}\right) \left(\frac{2}{\Delta}\right)^2 \right] \;.
\end{align}
For large \(N\) the binomial coefficient tends to a Gaussian and the normalized spin variable \(m_i = M_i/N\ \in [1,-1]\) becomes a continuous variable
\begin{align}
\label{eq:partition_binom}
 Z =& \sum_{\{M_{i}\}} A \exp \left[\frac{N}{2} \sum_{i,j}^L J_{i,j} m_i m_j - N \sum_{i=1}^L \frac{m_i^2}{2} \right]
\end{align}
with
\begin{align}
\label{eq:partition_binom_defs}
A =  \left(\frac{2^{2N+1}}{\pi N}  \right)^{\frac{L}{2}} \;,\quad J_{i,j} = \frac{1}{2 \beta} \tilde{K}_{\beta}\left(|j-i|\frac{2}{\Delta}\right) \left(\frac{2}{\Delta}\right)^2
\end{align}
and \(\tilde{K}_{\beta}(\tau) = N K_{\beta}(\tau)\). The form in (\ref{eq:partition_binom}) shows a one-dimensional continuous spin model with long-range interaction, where the states \(m_i \approx 0\) are favored due to the quadratic term. One can extract the \(N\)-dependence from this model, even without having to solve it in detail. Without the last term in (\ref{eq:partition_binom}), the factor \(N\) in the exponential would simply scales the coupling. As a consequence, the critical coupling would scaled as \(\tilde{\alpha}_c \propto N^{-1}\). In the following, we will therefore use temporarily the replacement
\begin{align}
  \label{eq:alpha_tilde_tilde}
  \tilde{\tilde{\alpha}} = N \tilde{\alpha}\;.
\end{align}
Effectively, this replacement is only introduced for bookkeeping the \(N\)-dependence in the following argumentation.

The effect of the last term in (\ref{eq:partition_binom}) can understood by its influence on the corresponding Ginzburg-Landau functional (GLF) \cite{Aharony,Luijten1997,Luijten1996}
\begin{align}
\label{eq:GLF}
\mathcal{F}[m(\tau)] &= \int \mathrm{d} \tau \Bigg[ \frac{r}{2}  m^2 - \frac{N}{2} m^2 + \frac{u}{4} m^4 + c\int \mathrm{d} \tau'  \frac{m(\tau)m(\tau')}{|\tau-\tau'|^{1+s}} \Bigg]
\end{align}
where
\begin{align*}
Z =& \int \mathcal{D} \left[m(\tau)\right] \exp \left[\mathcal{F}\left[m(\tau)\right] \right].
\end{align*}
The first and third terms are the common second and fourth order terms of the GLF and the last term represents the long range interaction in an integral form. The parameters \(r,u,c\) are all functions of \(s,\Delta,\tilde{\tilde{\alpha}}\), but do not dependent on \(N\) in an explicit manner. The local term in (\ref{eq:partition_binom}) now enters the functional as other quadratic term \(N m^2/2 \), that does \emph{not} depend neither on the coupling \(\tilde{\tilde{\alpha}}\) nor on other system parameters, but depends linearly on the number of spins \(N\). The parameter \(r\) can always shifted in such a way, that the phase transition occurs at \(r=0\). In the case where the \(N m^2/2\)-term is neglected, this parameter must then have the form \(r = b(s,\Delta) \left(\tilde{\tilde{\alpha}}_c - a(s,\Delta)\right)\), with unknown functions \(a\) and \(b\). In the full model with included \(N m^2/2\)-term, the equation
\begin{align}
\label{eq:determ_alpha}
r - N &= b(s,\Delta) \left(\tilde{\tilde{\alpha}}_c - a(s,\Delta)\right) - N = 0 \nonumber \\
\end{align}
determines the transition point \(\tilde{\alpha}_c\).

Even the offset function \(b(s,\Delta)\) can determined. Since the spin variables in (\ref{eq:partition_binom}) are continuous and independent of \(N\), a saddle point integration becomes exact in the limit \(N \rightarrow \infty\) and the onset of the magnetization of the most probable state gives phase transition point for \(N \rightarrow \infty\)
\begin{align}
\label{eq:b_pred}
\sum_{j} m_j J_{i,j} - m_i = 0 \; \overset{m \rightarrow 0^{+}}{\Rightarrow} \;  \tilde{\alpha} = \frac{s \Delta}{2 \omega_c} \mathrel{\mathop=}: b(s,\Delta)^{-1}\;.
\end{align}
Instead of the sum on the left-hand side, the integral over the Kernel \(\tilde{K}_{\beta}(\tau)/(\beta\Delta)\) is used (cf. Appendix~\ref{sec:mean-field_solution}) and the homogeneity of the spin variables at equilibrium \(m=m_k\) is assumed. With (\ref{eq:alpha_tilde_tilde}), (\ref{eq:determ_alpha}) and (\ref{eq:b_pred}), 
we arrive at the following prediction of the
asymptotic $N$, $s$, $\Delta$, and $\omega_c$ dependence of the 
critical coupling strength 
\begin{align}
\label{eq:alpha_c_pred}
\alpha_c\simeq \frac{1}{N}\cdot\frac{s\Delta}{2\omega_c}
+\frac{a(s,\Delta)}{N^2}
\end{align}

\subsection{The critical point: QMC results}
\label{secQMC}

For a sub-Ohmic bath ($0<s<1$) one expects a second order phase transition 
characterized by the following scaling laws ($N$, $s$,
$\Delta$, and $\omega_c$ fixed):
\begin{eqnarray}
m(\alpha=\alpha_c,\beta)&\propto&\beta^{-x/\nu}\\
\chi(\alpha=\alpha_c,\beta)&\propto&\beta^{\gamma/\nu}\label{chiscale}\\
Q(\alpha,\beta)&\sim &\tilde{Q}(\beta^{1/\nu}(\alpha-\alpha_c)/\alpha_c)
\label{Qscale}
\end{eqnarray}
At zero temperature (i.e. $\beta\to\infty$), then
$m(\alpha>\alpha_c,T=0)\propto(\alpha-\alpha_c)^x$
and $\chi(\alpha,T=0)\propto(\alpha_c-\alpha)^{-\gamma}$
holds. The critical exponents $x$ (usually denoted as $\beta$,
the order parameter exponent, which we changed to avoid confusion 
with the inverse temperature $\beta=1/T$), $\nu$ and $\gamma$ 
obey the scaling relation $\gamma=\nu-2 x$ and are
expected to be independent of the tunneling strength $\Delta$ 
and the cut-off frequency $\omega_c$, for which reason we
fix both to $\Delta=0.1$ and $\omega_c=1$ in most calculations.
The dependence of $\gamma$ and $\nu$ on $s$ is known for
$N=1$ [\onlinecite{Winter2009,Alvermann2009}], but not for general $N$.

\begin{figure}[t]
\includegraphics[width=0.48\textwidth]{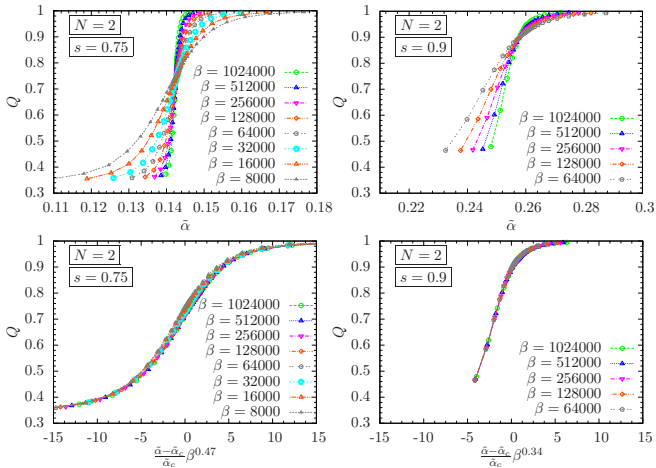}
\caption{\label{fig:4thorder_N2}
(Top row) Fourth order cumulant \(Q\) versus the rescaled bath coupling 
\(\tilde{\alpha}\) for \(s=0.75\) (left) and \(s=0.9\) (right) for \(N=2\). (Bottom row) Scaling plots of \(Q\) for the data in the top row for \(s=0.75\) (left) and \(s=0.9\) (right). The best data collapse is obtained
for $1/\nu=0.47$ for $s=0.75$ and $1/\nu=0.34$ for $s=0.9$.
The other parameters are \(\Delta = 0.1\) and \(\omega_c=1\).}
\end{figure}

\begin{figure}
\includegraphics[width=0.48\textwidth]{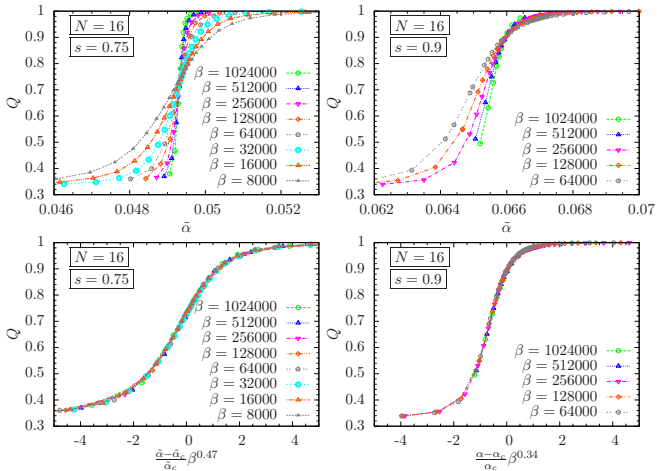}
\caption{\label{fig:figure2}
The same as Fig.\ \ref{fig:4thorder_N2}, but for $N=16$ instead of
$N=2$. The scaling plot in the lower row yields the best data collapse
for a different critical rescaled bath coupling but the same
values for the exponent $1/\nu=0.47$ for $s=0.75$ and $1/\nu=0.34$ for $s=0.9$.}
\end{figure}

We performed large scale QMC simulations simulations for \(N=1,\dots,128\) and \(\beta=2000, \dots , 1024000\) using a continuous imaginary time
algorithm based on the worldline representation (see subsection \ref{sec:pathintegral}) and [\onlinecite{Winter2009}]).
To determine the critical point $\alpha_c$ it is most 
convenient to use the relation (\ref{Qscale}): For fixed
$N$, $s$, $\Delta$, and $\omega_c$ the quantity $Q$
is at the critical point $\alpha=\alpha_c$ asymptotically 
independent of $\beta$, which can be used to locate the critical 
point. This is demonstrated for $s=0.75$
and $s=0.9$ in Fig.\ \ref{fig:4thorder_N2} and Fig.\ \ref{fig:figure2} for $N=2$ and $N=16$ respectively.

\begin{figure}
\includegraphics[width=0.4\textwidth]{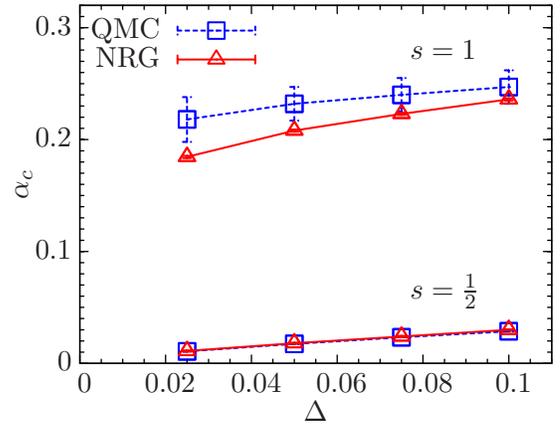}\\
\caption{\label{fig:comp_QMC-NRG} 
Comparison of QMC and NRG predictions for the critical bath coupling
$\alpha_c$ as a function of the tunneling
amplitude for $N=2$, $s=1/2$ and $s=1$. Squares are our QMC estimates 
and triangles the prediction of the NRG calculation of Ref.~[\onlinecite{LeHur2010}].}
\end{figure}

As a check for accuracy we compare the estimates for $\alpha_c$
that we obtain in this way for $N=2$ with the predictions of the 
numerical renormalization group (NRG) calculation of Ref.~[\onlinecite{LeHur2010}], which is shown in Fig.~\ref{fig:comp_QMC-NRG}.
The agreement is very well in the sub-Ohmic regime for \(s=1/2\) but differs slightly for an Ohmic spectrum (\(s=1\)). The phase transition of the latter case is known to be notorious difficult to investigate with Monte-Carlo simulations, because it belongs to the Kosterlitz-Thouless universality class. Throughout this paper, we set \(\Delta = 0.1\) and \(\omega_c = 1\), for which the two independent methods coincide within 5\% for the Ohmic case (which is compatible with the error bar) and much less in the sub-Ohmic regime.

\begin{figure}
\includegraphics[width=0.48\textwidth]{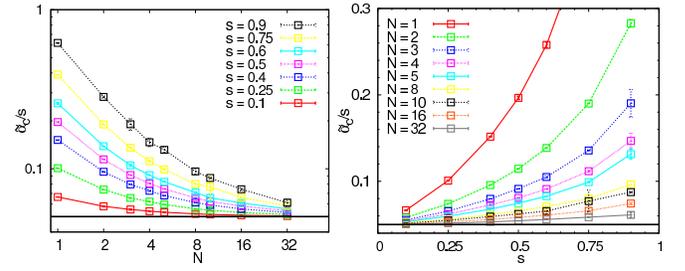}
\caption{\label{fig:alpha_c_vs_s+N} (Left) Scaled critical coupling \(\tilde{\alpha}_c/s\) versus the number of spins for several bath exponents \(s=0.1, \dots, 0.9\) (from bottom to top). The solid line indicates the limiting value of \(\frac{\Delta}{2 \omega_c}\) for \(N \rightarrow \infty\). (Right) \(\tilde{\alpha}_c/s\) versus the bath exponent \(s\). The parameters are \(\Delta = 0.1\) and \(\omega_c=1\).}
\end{figure}

\begin{figure}
\includegraphics[width=0.45\textwidth]{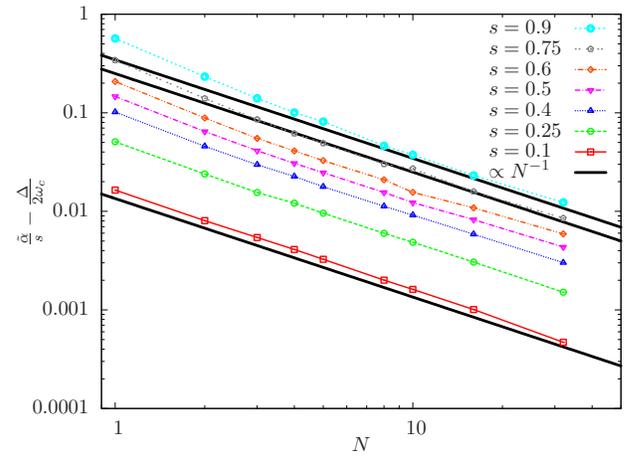}\\
\caption{\label{fig:alpha_c_vs_N} 
Check of the prediction (\ref{eq:alpha_c_pred}) for
the rescaled critical bath coupling
$\tilde{\alpha}_c/s - \Delta/(2 \omega_c)\propto1/N$
for several bath exponents \(s=0.1, \dots, 0.9\) (from bottom to top). 
The solid lines are a guide for the eyes.}
\end{figure}

Next we determined the critical bath coupling strength for
different values of $N$ and $s$, the result is shown in Fig.~\ref{fig:alpha_c_vs_s+N}.
For increasing $N$ the rescaled critical bath coupling strength appears to approach an $s$-dependent constant, which means that asymptotically (for
large $N$) $\alpha_c$ would decrease as $1/N$. In the last subsection
we derived a more precise prediction for the asymptotic behavior of $\alpha_c$ in Eq. (\ref{eq:alpha_c_pred}):
\begin{equation}
\alpha_c\simeq \frac{1}{N}\cdot\frac{s\Delta}{2\omega_c}
+\frac{a(s,\Delta)}{N^2}\nonumber
\end{equation}
This prediction is checked in Fig.~\ref{fig:alpha_c_vs_N}. The simulation results show a quite good agreement with Eqs.~(\ref{eq:alpha_c_pred}) and (\ref{eq:b_pred}) even at comparatively small \(N\). 

According to the theoretical considerations of the last subsection, 
the universality classes of the MSBM model for general, finite $N<\infty$ should be identical with the universality class of the single-spin-boson model. More precisely, one expects a Gaussian fixed point for
\(s<1/2\) (i.e.\ $x=1/2$, $\gamma=1$, and $\nu=1/s$, see
[\onlinecite{Fisher1972,Luijten1997}]), non-trivial exponents
for \(1/2 < s \leq 1\) and no phase transition above \(s=1\). 
Since the quantum phase transition of the single-spin-boson model
is well described by the zero temperature phase transition in
the mean-field model (\ref{eq:action_MFT}) (see also [\onlinecite{HouTong2010}]),
one expects in the regime $s<1/2$ mean-field exponents also for the 
MSBM with $N>1$. We confirmed with our QMC simulation that
the classical exponents $x=1/2$, $\gamma=1$, and $\nu=1/s$
for $s<1/2$ are indeed independent of $N$ (data not shown)
and focus here on the more interesting, the non-classical 
regime $1/2<s\leq1$, and show results for two explicit values: 
\(s=0.75\) and \(s=0.9\). 

\begin{figure}
\includegraphics[width=0.45\textwidth]{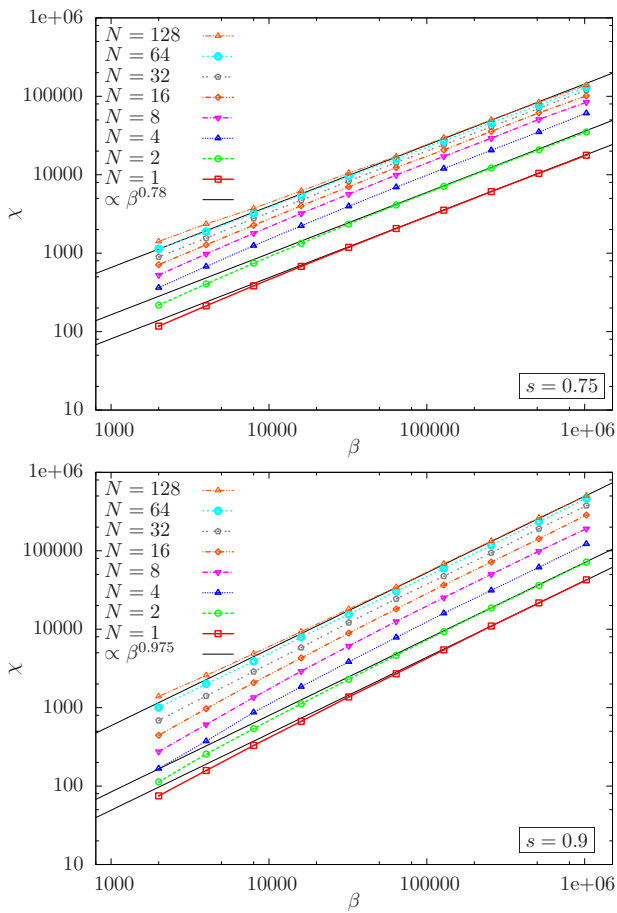}
\caption{\label{fig:chi_c_vs_beta} Susceptibility at the critical point versus the inverse temperature \(\beta\) for \(s=0.75\) (top) and  \(s=0.9\) (bottom) for different $N$. The straight line represents the asymptotic behavior $\chi\propto\beta^{\gamma/\nu}$ with $\gamma/\nu=0.78$ for 
$s=0.75$ and $\gamma/\nu=0.975$ for $s=0.9$.}
\end{figure}

Fig.\ \ref{fig:4thorder_N2} and \ref{fig:figure2} show scaling plots
of $Q$ according to the scaling relation (\ref{Qscale}). Since 
corrections to scaling increase noticeably
if \(s\) approaches \(1\), we restricted the scaling plot to 
\(\beta \ge 64000\) for \(s=0.9\),
whereas for \(s=0.75\) the data for \(\beta \ge 8000\) has been 
included for scaling. The data collapse 
is very good for the exponent values $1/\nu=0.47$ and $1/\nu=0.34$ for
$s=0.75$ and $s=0.9$, respectively, for both $N=2$ and $N=16$.
These estimates for $\nu$ also agree with those that we obtained for 
$N=1$ (data not shown, c.f.~[\onlinecite{Winter2009}]) and therefore support
our prediction of the last subsection that the universality class
of the MSBM is independent of $N$ for finite $N$.

We obtain a second independent exponent, namely $\gamma$, from the behavior of the susceptibility $\chi$ (\ref{chiscale}) at the critical point. The data are shown in Fig.~\ref{fig:chi_c_vs_beta} in a log-log plot. For both bath exponents the asymptotic slope of the susceptibility \(\chi \simeq \beta^{\gamma/\nu}\) stays unchanged as the number of spins is increased, which confirms that the critical exponent is $\gamma/\nu=0.78$ for $s=0.75$ and $\gamma/\nu=0.975$ for $s=0.9$ is universal for any value of $N$. 

The pre-asymptotic behavior ($\beta\ll 10^4$) displays small systematic deviations from the straight line. For small $N$ the asymptotic straight line is approached from below indicating corrections to scaling. For large $N$ it is approached from above indicating a crossover from the $N=\infty$ mean-field critical behavior to the asymptotic single-spin-boson behavior $N=1$. 

\begin{figure}[b]
\includegraphics[width=0.48\textwidth]{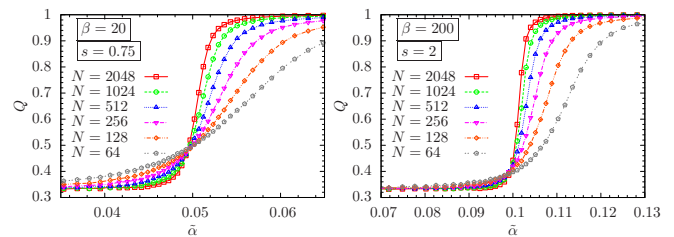}
\caption{\label{fig:Q_vs_slpha_for_N}Fourth order cumulant \(Q\) versus the rescaled critical coupling \(\tilde{\alpha}\) for several number of spin \(N\) for the parameters \(\beta = 20\) and \(s=0.75\)  (left) and \(\beta = 200\) and \(s=2\) (right).}
\end{figure}

Finally we study the $N\to\infty$ limit of the MSBM model, for which 
we expect the mean-field theory described in section \ref{sec:finiteT}
to be exact. Since the latter has a phase transition also at non-vanishing
temperature (\ref{eq:T_alpha_c_MFT}) we performed extensive QMC simulations 
for fixed temperatures (fixed $\beta$) and \(\Delta = 0.1\), \(\omega_c=1\). Fig.~\ref{fig:Q_vs_slpha_for_N} shows the fourth order cumulant \(Q=\langle m^2 \rangle^2/\langle m^4 \rangle\) versus the control parameter (i.e \(\tilde{\alpha}\) in our case) for multiple numbers of spins \(N\) for two exemplary cases. One can clearly see that the curves intersect in a single point, which determines the phase transition very similar the previous zero temperature analysis, where \(N\) was held fix.

\begin{figure}
\includegraphics[width=0.45\textwidth]{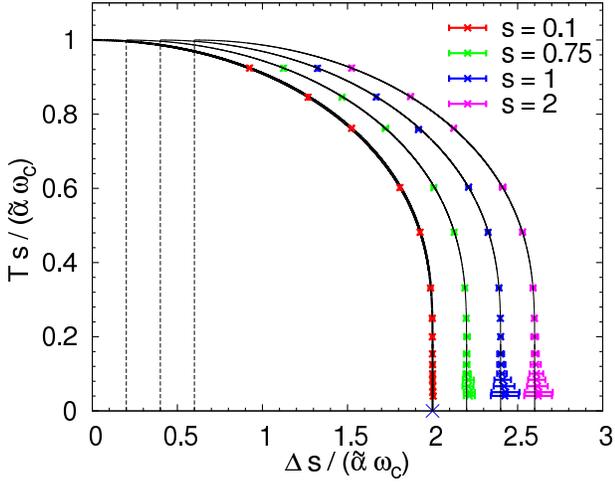}
\caption{\label{fig:finiteT} Temperature versus tunneling amplitude \(\Delta\) (both in unit of \(\omega_c \tilde{\alpha}/s\)) for different bath exponents \(s\). The thick black curve shows the analytic mean-field result (\ref{eq:alpha_c_MFT}) and the data points are obtained by QMC in the limit of \(N \rightarrow \infty\) (see text). The results for \(s = 0.75, 1, 2\) have been shifted by 0.2, 0.4 and 0.6 respectively (indicated by the dashed lines) for better visibility. The blue cross is the zero-temperature limit (\ref{eq:MF_alpha_c}). The fixed parameters are \(\Delta = 0.1\) and \(\omega_c=1\).}
\end{figure}

The resulting temperature versus tunneling plot (both in units of the scaled coupling \(\tilde{J} = \frac{\omega_c \tilde{\alpha}}{s}\)) is shown in Fig.~\ref{fig:finiteT}. All points coincide to a single master curve (cf. Eq.~(\ref{eq:T_alpha_c_MFT})). One can see an excellent agreement with the expected transition line. All curves tend to the zero temperature limit of (\ref{eq:alpha_c_MFT})
\begin{align}
\label{eq:MF_alpha_c}
\Delta_c = \frac{2\tilde{\alpha} \omega_c}{s}\;,
\end{align}
which is marked as a blue cross in Fig.~\ref{fig:finiteT}. At low temperatures and bath exponents \(s \geq 1\), an increase of the size of the statistical error (for constant computational effort) is clearly visible. This is an indication of the onset of the crossover regime between mean-field universality at finite \(T\) and \(N \rightarrow \infty\) and the non-classical critical behavior governed by the quantum critical point at \(T=0\). For the same reason one observes the asymptotic behavior at the 
quantum critical point for large $N$ only for large $\beta$, i.e.\ low temperatures $T$.

\section{Spin-spin correlation}
\label{sec:MSBM_corrfunc}

In this section, we investigate the effective interaction between the two-level systems mediated by the common bath by calculating the spin-spin correlation functions of the MSBM via QMC. It is defined as 
\begin{align}
\label{eq:corrfunc_def}
\langle \sigma_1^z \sigma_2^z\rangle = \frac{1}{Z}\operatorname{Tr}\left(\sigma_1^z \sigma_2^z \exp(-\beta H) \right)\;,
\end{align}
which is easily accessible by the quantum Monte Carlo algorithm, since the operator \(\sigma_1^z \sigma_2^z\) is diagonal in the used representation.

\begin{figure}
\includegraphics[width=0.45\textwidth]{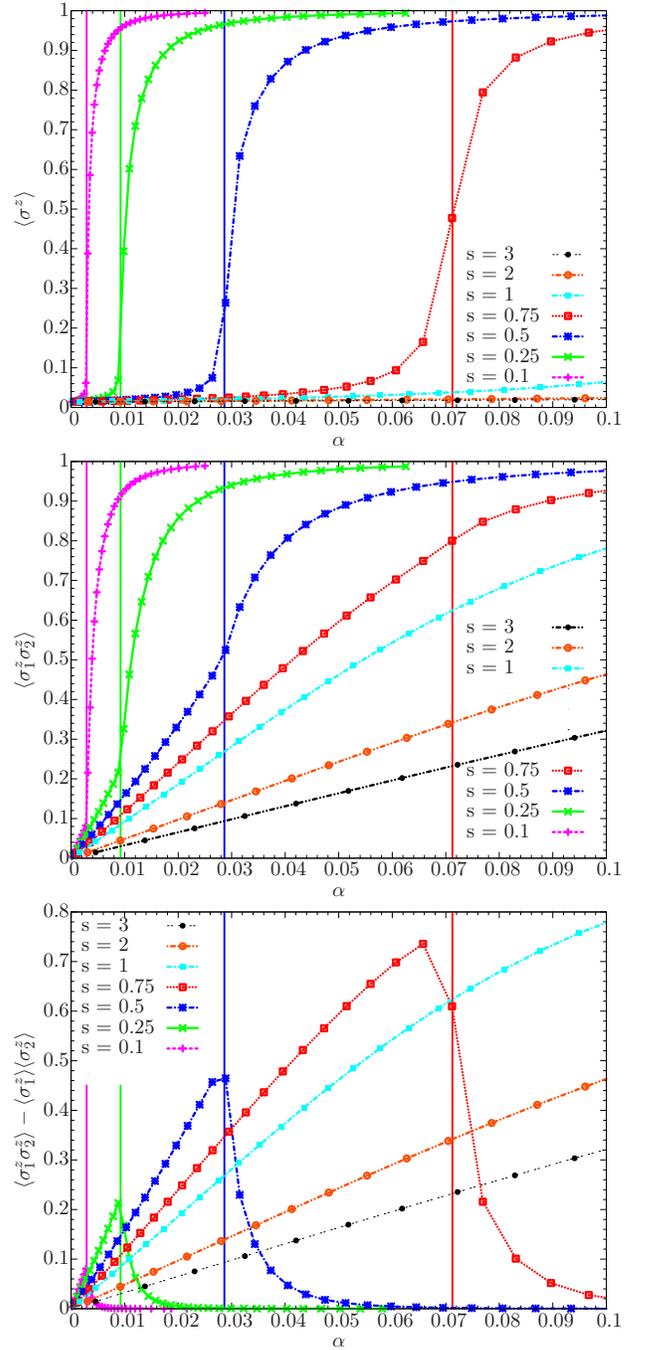}
\caption{\label{fig:obs_vs_alpha_sall} Order parameter \(\langle \sigma^z \rangle\) (top), spin-spin correlation function \(\langle \sigma_1^z \sigma_2^z \rangle\) (middle) and connected correlation function \(C_{1,2} = \langle \sigma_1^z \sigma_2^z\rangle - \langle \sigma_1^z \rangle \langle \sigma_2^z\rangle\) versus the coupling strength \(\alpha\) for various bath exponents \(s\). The vertical lines corresponds to the phase transition points.  The inverse temperature is \(\beta = 32000\).}
\end{figure}

In Fig.~\ref{fig:obs_vs_alpha_sall} the order parameter \(\langle \sigma^z \rangle = \langle \sigma_1^z \rangle = \langle \sigma_2^z \rangle\), the correlation function (\ref{eq:corrfunc_def}) and the connected correlation function \(C_{1,2} = \langle \sigma_1^z \sigma_2^z\rangle - \langle \sigma_1^z \rangle \langle \sigma_2^z\rangle\) are plotted. The vertical lines display the phase transition points \(\alpha_c\) with respect to the different bath exponents  \(s\). As already shown in Sec.~\ref{sec:QPT_MSBM}, \(\alpha_c\) is shifted towards larger coupling strengths as \(s\) is increased. At zero temperature ($\beta=\infty$) the order parameter
$\langle \sigma^z \rangle$ is zero in the delocalized phase 
($\alpha<\alpha_c$) and increases with an algebraic singularity
$\propto(\alpha-\alpha_c)^x$ in the localized phase ($\alpha > \alpha_c$).
For finite $\beta$ this sharp transition is smoothed as is visible
in Fig.~\ref{fig:obs_vs_alpha_sall}(a).

The correlation function \(\langle \sigma_1^z \sigma_2^z\rangle\) in Fig.~\ref{fig:obs_vs_alpha_sall}(b) is non-zero even in the delocalized phase \(\alpha < \alpha_c\) due to the effective ferromagnetic interaction
mediated by the common bath. We will quantify this behavior for small coupling strengths at the end of this section. For \(\alpha > \alpha_c\) the non-zero order parameter (Fig.~\ref{fig:obs_vs_alpha_sall}(a)) superposes with the correlation function. This leads to a kink at \(\alpha_c\) , that is clearly pronounced, if \(s\) is small and smeared out, if \(s\) becomes larger.

In Fig.~\ref{fig:obs_vs_alpha_sall}(c), the connected correlation function \(C_{1,2}\) is shown. This property displays the fluctuation around the mean value. One can see, that this function grows until the phase transition point is reached and decreases rapidly to zero after passing it. The larger the bath exponent is, the higher the maximum of the curve is until the phase transition to the localized phase suppresses the fluctuations. For \(s>1\), where no phase transition takes place, \(C_{1,2}\) is monotonically increasing and saturates at \(C_{1,2}=1\) for large \(\alpha\). In this case, the spins are strongly correlated, but not able to perform a localization. In the path integral representation, the two worldlines arranges themselves in synchronized way, where their kinks occur at nearly equal times, separating identically orientated worldline segments from each other.

\begin{figure}
\includegraphics[width=0.45\textwidth]{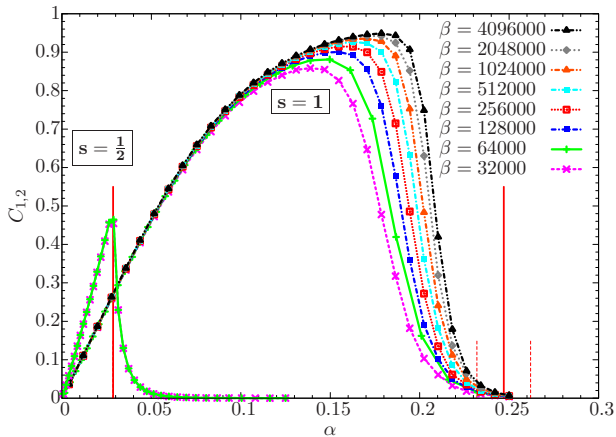}
\caption{\label{fig:comp_QMC-NRG_entanglement} Connected correlation function \(C_{1,2}\) versus the coupling strength for the Ohmic case \(s=1\), where data from \(\beta = 32000\) to \(\beta = 4096000\) is shown and for \(s=\frac{1}{2}\) with \(\beta = 32000\) and \(\beta=64000\). The vertical solid line marks the phase transition point and the dashed lines its error.}
\end{figure}

Fig.~\ref{fig:comp_QMC-NRG_entanglement} displays \(C_{1,2}\) for a larger range of \(\alpha\), where the  behavior for an Ohmic bath spectrum is also covered. Plotted are curves for different inverse temperatures. One observes  that even the largest system  size with \(\beta=4096000\) still shows strong finite $\beta$ effects in contrast to systems in the sub-Ohmic regime, where e.g. for \(s=1/2\) the curves for \(\beta=32000\) and \(\beta=64000\) collapse already quite well. The origin of this slow convergence to the 
infinite $\beta$ limit can be understood by the particular critical behavior at \(s=1\). At this point, the kernel \(K_{\beta}(\tau)\) falls off in imaginary time with an inverse-square law. This kind of long-range interaction is known to produce a Kosterlitz-Thouless transition, where the spin-spin correlation function decays logarithmically with the distance in imaginary time\cite{Scalapino1981,Messingfeld}. Therefore, finite size effects survive over many decades causing much larger deviations in the numerical determined properties. The determination of the phase transition point is performed by an extrapolation of the fourth order cumulant \(Q=\langle m^2 \rangle^2/\langle m^4 \rangle\) to infinite $\beta$ and is for $s=1$ accompanied by a much larger error bar then for the sub-Ohmic regime ($s<1$).

\begin{figure}
\includegraphics[width=0.45\textwidth]{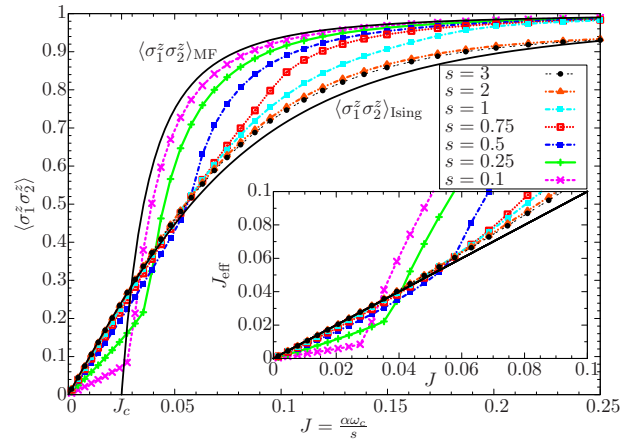}
\caption{\label{fig:corrfunc_full2} Correlation function of two spins in a common bath versus the rescaled coupling parameter \(J=\alpha \omega_c /s\) for various bath exponents \(s\). The solid lines are the analytic results of the pure Ising coupling or the mean-field interaction respectively (see text). The inset shows the effective direct interaction \(J_{\mathrm{eff}}\) versus \(J\) for small couplings strengths (\(\alpha=0 \dots 0.1\)). The solid line is the identity \(J_{\mathrm{eff}} = J\).}
\end{figure}

The observed behavior of the connected correlation function 
is reminiscent of the entanglement entropy
\begin{align}
\mathcal{E}=-\operatorname{Tr} \left[\rho_s \log_2 \rho_s \right]\;,
\end{align}
with the reduced density matrix \(\rho_s= \operatorname{Tr}_B \rho\), where the bath degrees of freedom were traced out. The entanglement entropy was recently studied in Ref.~[\onlinecite{LeHur2010}], where Fig.~6 shows also a cusp for the sub-Ohmic case \(s=1/2\) whereas Fig.~5 therein shows a smooth maximum for the Ohmic spectrum (note that vanishing direct spin-spin-coupling $K=0$ corresponds to the case considered by us). 

We will now discuss some particular regimes of the model. In Ref.~[\onlinecite{LeHur2010}] it was shown, that a polaron transformation \(U=\exp \left(-\frac{1}{2} (\sigma_1^z + \sigma_2^z) \sum_k \frac{\lambda_k}{\omega_k} (b_k^{\dagger} - b_k) \right)\) applied to the two impurity spin-boson model will renormalize a bare (anti-ferromagnetic) Ising coupling term \(\frac{K}{4} \sigma_1^z \sigma_2^z\) to \(\left(\frac{K}{4} - \frac{\alpha \omega_c}{s}\right) \sigma_1^z \sigma_2^z\). Since we consider no bare direct coupling between the spins (\(K=0\)), it is convenient to plot the data against the rescaled coupling strength \(J=\alpha \omega_c / s\) to eliminate the global \(\propto s^{-1}\)-scaling of the bath induced spin-spin interaction, if one is comparing the behavior for different \(s\). In Fig.~\ref{fig:corrfunc_full2} the data of Fig.~\ref{fig:obs_vs_alpha_sall}(b) is plotted versus \(J\). By means of this plot, one can see, that the curves for high \(s>1\) tend to a master curve. This master curve corresponds to the analytic result of two non-dissipative and Ising-coupled spins driven by a transverse field
\begin{align}
\label{eq:Ham_direct}
H = \frac{\Delta}{2} \left(\sigma_1^x + \sigma_2^x\right) - J \sigma_1^z  \sigma_2^z\;.
\end{align}
This toy-model is exactly solvable and one could calculated the spin-spin correlation to be
\begin{align}
\label{eq:corrfunc_Ising}
\langle \sigma_1^z \sigma_2^z\rangle_{\mathrm{Ising}} = \frac{J}{\sqrt{J^2+\Delta^2}}
\end{align}
in the zero temperature limit.

For $s\to0$ the data are compatible with mean-field behavior as can be seen by solving the MSBM in mean-field approximation (c.f Eq.~\ref{eq:action_MFT}) for finite $N$. Within this approximation, the model shows a phase transition at \(J_c = \Delta /(2 N) \) as discussed in Sec.~\ref{sec:finiteT} and the correlation function, that is in this case simply the order parameter squared, reads
\begin{align}
\label{eq:corrfunc_MF}
\langle \sigma_1^z \sigma_2^z\rangle_{\mathrm{MF}} = \langle \sigma^z \rangle^2 = 
\begin{cases}
0 & \mbox{ if } J \leq J_c =\frac{\Delta}{2 N}\\
1-\frac{\Delta^2}{4 N^2 J^2} & \mbox{ if } J > J_c\;.
\end{cases}
\end{align}
The data for $\langle \sigma_1^z \sigma_2^z\rangle$ for the lowest bath exponent \(s=0.1\) are quite close to
the mean-field results (\ref{eq:corrfunc_MF}) in the localized phase.
In the delocalized phase $\langle \sigma_1^z \sigma_2^z\rangle$ does not vanish, in contrast to mean-field behavior, and also deviates from 
the behavior of the two coupled spins without bath (\ref{eq:corrfunc_Ising}).
To shed light on this weak coupling regime, we recast Eq.~(\ref{eq:corrfunc_Ising}) and define an effective direct coupling
\begin{align}
\label{eq:corrfunc_Ising_invert}
J_{\mathrm{eff}} = \frac{\Delta \, \langle \sigma_1^z \sigma_2^z\rangle}{\sqrt{1-\langle \sigma_1^z \sigma_2^z\rangle^2}}
\end{align}
that is a measure for the effective ferromagnetic coupling mediated by the common bath.

The inset of Fig.~\ref{fig:corrfunc_full2} shows the effective direct coupling (\ref{eq:corrfunc_Ising_invert}) versus the rescaled coupling \(J=\alpha \omega_c / s\). Again, one sees the approach of the curves for high \(s\) to the identity (\(J_{\mathrm{eff}} = J\)), whereas for small \(s\) the effective direct coupling develops significantly slower than \(J\), but still linear. The data imply that the ratio $J_{\mathrm{eff}}/J$ decrease to zero 
for $s\to0$, which is compatible with the mean-field approximation, since the results (\ref{eq:corrfunc_MF}) predicts \(J_{\mathrm{eff}}=0\) in the delocalized phase. For $s>1$ the ratio $J_{\mathrm{eff}}/J$ approaches one, the effective interaction mediated by the common bath is just
$J_{\mathrm{eff}}=J=\alpha\omega_c/s$. Note that even for $s\to\infty$,
i.e\ short-ranged interactions in the imaginary time direction,
the two-level systems are still ferromagnetically correlated due
to the effective bath interaction at equal imaginary times.
The $s\to\infty$ limit of the $(N=2)$-MSBM is thus simply
the pair of two-level system without
bath but with an ferromagnetic coupling $J=\alpha\omega_c/s$.

\begin{figure}
\includegraphics[width=0.48\textwidth]{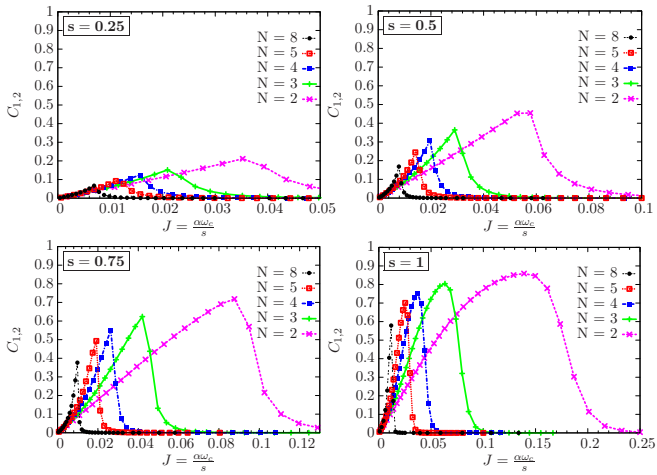}
\caption{\label{fig:corrfunc_full2_nimp} Correlation function of multiple spins in a common bath versus the coupling parameter \(J= \alpha \omega_c /s\) for bath exponents \(s=0.25,0.5,0.75\) and \(s=1\) at an inverse temperature of \(\beta=32000\)}
\end{figure}

Finally, we had a look on the correlations for more than two spins. Fig.~\ref{fig:corrfunc_full2_nimp} shows the connected correlation function for \(N=2\) to \(N=8\) for different bath exponents. One can see, that the maximum becomes lower and more sharp as the number of spins is increased. For larger \(N\) the maximum would be takes place at the mean-field prediction of \(J_c = \Delta/(2N)\) (\ref{eq:corrfunc_MF}). The initial slope for \(J \rightarrow 0\) is quite similar for all curves, but raises super-linearly for \(N>2\) instead of linear for \(N=2\), when the coupling is increased. That means, that the spins build up their cooperative polarization successively.

\section{Spatial separation}
\label{sec:spa_sep}
\begin{figure}
\includegraphics[width=0.45\textwidth]{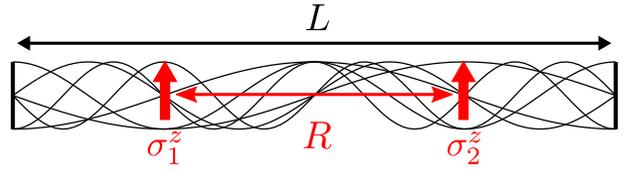}
\caption{\label{fig:sep_sep_moti} Sketch of two spins at a distance \(R\) embedded in a one-dimensional common bosonic bath. The bath is fully described by a complete set of harmonic modes, that are restricted to a box with linear size \(L\).}
\end{figure}

In this section we consider spatially separated two-level systems in a common bath. As originally proposed in Ref.~[\onlinecite{Stamp1998}] the two-level systems are now coupled to a phase-shifted polarization of the bosonic modes. The phase-shift arises from the time \(t=R/v\), that is needed propagate the information from one spin to another spin at a distance \(R\) with the propagation velocity \(v\). Such a situation arises for instance if the bosonic modes are represented by standing waves in a box with a linear size \(L\) and periodic boundary conditions (Fig.~\ref{fig:sep_sep_moti}). In this case, the bath is fully described by a complete set of harmonic modes \(\mathbf{k}= \mathbf{n} \! \cdot \! 2 \pi /L\) with ${\mathbf n}=(n_1,\ldots,n_d)$, and $n_i=1,2,3,\ldots$ for $i=1,\ldots,d$ and $d$ the dimension of the box. After performing \(L\rightarrow \infty\), this approach is basically a continuous version of phonon induced interaction between localized electrons \cite{Mahan}. The same distance-dependent interaction has also used for two harmonic oscillators in a common bath instead of two spins \cite{Zell2009}.

\subsection{The dissipative spin array}
\label{sec:diss_ary_spa_sep}
The generalization of interaction term of the MSBM (\ref{HI_MSBM}) to a system, where the \(N\) spins are located at the positions \(\mathbf{r_m}\) is given by \cite{McCutcheon2010,Carrbook}
\begin{align}
\label{eq:Ham_SBM_HI_spasep}
H_{I} =\sum_{m=1}^{N}  \frac{1}{2}\sum_{\kk} \lambda_\kk \left( a^{\dagger}_\kk e^{i \kk \cdot \mathbf{\rr_m}} + a_{\kk} e^{-i \kk \cdot \mathbf{\rr_m}}\right) \sigma_m^z\;,
\end{align}
Without loss of generality, the coefficients \(\lambda_k\) are assumed to be real numbers, because a complex phase carries no information. In contrast to the previous model without any relation to spatial variables, the indices have now the meaning of actual wavenumbers rather than a simple counter for the frequencies. In principle, this would involve a dispersion relation \(\omega \rightarrow \omega(\kk)\), which would also affect the couplings \(\lambda_k \rightarrow \lambda_{\omega(\kk)}\). Since we are only considering modes propagating in vacuum (\(\omega = v |\kk|\)), we stick to the sloppy notation of Eq.~(\ref{eq:Ham_SBM_HI_spasep}).

One can show (cf. Appendix~\ref{sec:deriv}) that the dependence on \(R=|\mathbf{R}| = |\mathbf{r_m} - \mathbf{r_{m'}}|\) can be completely integrated  into the spectral function
\begin{align}
\label{eq:spectral_spasep_sum} 
J(\omega,\mathbf{R})  &= \pi \sum_{\kk>0} \lambda_{\kk}^2\cos \left(\kk \!\cdot\! \mathbf{R} \right) \delta(\omega - \omega_{\kk})
\end{align}
such that the form of the kernel (\ref{eq:kernel}) remains unchanged. The sum in (\ref{eq:spectral_spasep_sum}) can be replaced by an integral in the limit \(L \rightarrow \infty\) and the couplings are assumed to follow a power-law function \(\lambda_{k_i} \propto k_i^{(2+s-d)/2}\) (\(i=1,\dots,d\)).

Absorbing all constants into the bath coupling strength  \(\alpha\) leads to the form
\begin{align}
\label{eq:spectral_spasep}
J(\omega,\mathbf{R}) &= 2 \pi \alpha \omega^s \omega_c^{1-s}f^d\left( \bar{R} \right)\;,
\end{align}
where we have introduced the scaled distance \(\bar{R}=R \omega_c / v\). The dependence on the spatial separation is thus absorbed into a single function \(f^d(\bar{R})\). For the different dimensions \(d=1,2,3\) of the bath, this function is
\begin{align}
f^d(\bar{R}) = 
\begin{cases}
\cos(\bar{R}) & d=1 \\
\operatorname{J_0}(\bar{R}) & d=2 \\
\sin(\bar{R})/\bar{R} & d=3 
\end{cases}\;,  \label{eq:f_dD}
\end{align}
where \(\operatorname{J_0}(\bar{R})\) is the Bessel function of the first kind.

\begin{figure}
\includegraphics[width=0.45\textwidth]{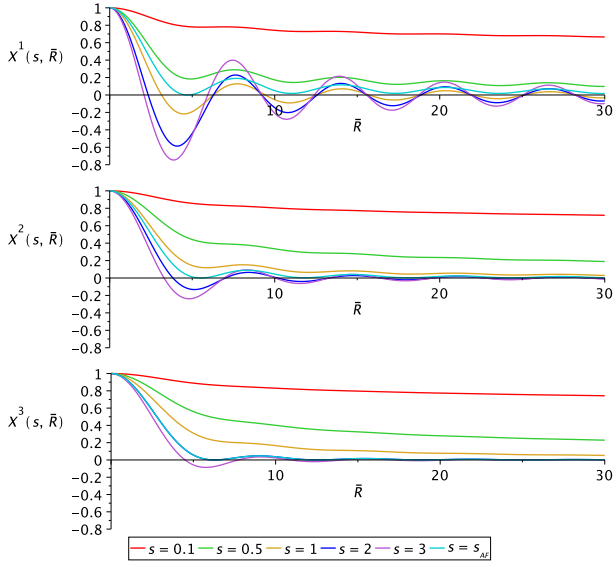}
\caption{\label{fig:X_dD}\(X^d(s,\bar{R})\) in one, two and three dimensions for various \(s\). For \(d=3\), the curve for \(s=2\) coincides with one for \(s=s_{AF}\).}
\end{figure}

Let us focus on the separation dependent part and consider the zero tunneling limit \(\Delta = 0\). The approach of Sec.~\ref{sec:zerotunnelinglimit} can easily generalized and the resulting partition function corresponds to the one of a long-range Ising model
\begin{align}
Z = \sum_{\{\sigma_1, \dots, \sigma_N\}} \exp \left( \frac{\beta}{2} \sum_{m=1}^N \sum_{m'=1}^N \sigma_m \sigma_{m'} \frac{\alpha \omega_c }{s} X^d(s,\frac{\omega_c}{v}\left|\rr_m-\rr_{m'}\right| \right)\;,
\end{align}
where the Ising interaction is defined as
\begin{align}
\label{eq:X_dD}
X^d(s,\bar{R}) = \int_0^{1} s x^{s-1} f^d(x \bar{R})\mathrm{d}x\;.
\end{align}

For every dimension \(d=1,2,3\), these functions reproduce the MSBM case from Sec.~\ref{sec:zerotunnelinglimit}, if the scaled distance \(\bar{R} \rightarrow 0\). In Fig.~\ref{fig:X_dD}, \(X^d(s,\bar{R})\) is plotted for different \(s\) in one, two and three dimensions. For small \(s\) they decay monotonously, whereas for higher \(s\) oscillations become dominant. The asymptotic behavior of the former case can be determined by recasting (\ref{eq:X_dD}) to \(X^d(s,\bar{R}) = \bar{R}^{-s} I^d(s,\bar{R})\) and noting that the remaining integral \(I^d(s,\bar{R}) = \int_0^{\bar{R}} s x^{s-1} f^d(x)\mathrm{d}x\) converges to a finite (non-zero) value as \(\bar{R} \rightarrow \infty\). It turns out, that
\begin{align}
\label{eq:X_dD_asymp}
X^d(s,\bar{R}) \simeq \bar{R}^{-s}
\end{align}
for \(0 < s < (d+1)/2\).

Above this value \(s > (d+1)/2\) this analysis breaks down and one should rather focus on the envelop function of \(X^d(s,\bar{R})\). One way is to use the form (e.g in \(d=1\)) \(X^1(s,\bar{R}) =  \bar{R}^{-1} \int_0^{\bar{R}} s \left(x/\bar{R}\right)^{s-1} \cos(x)\mathrm{d}x\) and analyzing the asymptotic behavior of the maxima at \(\bar{R}_n = 2 \pi n + \pi/2\) for large \(n\). One obtains the \(s\)-independent power-law decay
\begin{align}
\label{eq:X_dD_asymp_above}
X^d(s,\bar{R}) \simeq \bar{R}^{-(d+1)/2}\;,
\end{align}
for the envelop function, if \(s > (d+1)/2\).

In addition to the asymptotic behavior another upper bound for the bath exponent is important that also depends on the dimension of the bath. We define \(s_{AF}^d\) such that for all \(s > s_{AF}^d\) the coupling interaction \(X^d(s,\bar{R})\) is not restricted to positive values for all distances any more. Intervals of the distance \(\bar{R}\) occurs, for which anti-ferromagnetic interactions occur. The onsets of anti-ferromagnetism are in this sense \(s_{AF}^1=0.6923(9)\), \(s_{AF}^2=1.3545(2)\) and \(s_{AF}^3=2.001(2)\). For bath exponents smaller than these values the spin-spin interactions are always ferromagnetic and decay asymptotically as discussed above.

The lower bound for feasible values of the bath exponents follows from
the following consideration:
It is known, that a \(D\)-dimensional Ising model with algebraic decaying long-range interaction does only have a non-diverging energy per spin, if the decay is faster than \(1/R^{D}\) (see Ref.~[\onlinecite{Luijten1997}] and references therein). That means for example, that in the present case the one-dimensional infinite spin chain in a \(d\)-dimensional bosonic bath can only exists, if \(s \geq 1\).

\begin{figure}[t]
\includegraphics[width=0.33\textwidth]{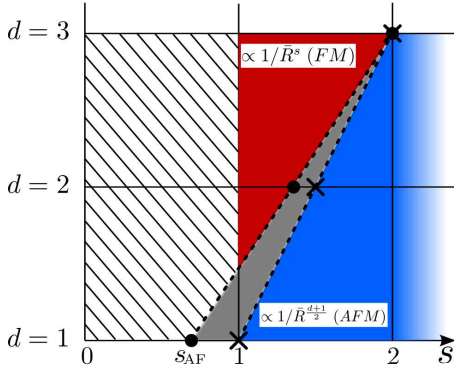}
\caption{\label{fig:classification} Sketch of the phases of the dissipative spin array in the zero tunneling limit for a one, two and three-dimensional bath. (white lined) A decay slower than \(\propto R^{-1}\) is thermodynamically not possible in an infinite chain. (red) Ferromagnetic interaction (FM) with \(\propto R^{-s}\) decay. (blue) Anti-ferromagnetic interaction (AFM) with \(s\)-independent \(\propto R^{-\frac{d+1}{2}}\) decay. (gray) Intermediate regime between (FM) and (AFM).} 
\end{figure}

These considerations are summarized in Fig.~\ref{fig:classification}. For \(1 \leq s \leq s_{AF}^d\), the system has strict ferromagnetic interactions and 
is equivalent to the classical Ising spin chain with algebraically decaying long-ranged interactions \cite{Luijten1997}. This regime exists in the range of bath exponents \(1 < s < 2\) in the case of a three-dimensional bath and vanishes for \(d=1\). For \(s \geq s_{AF}\) the behavior depends on the details the lattice spacings \(|\rr_m - \rr_{m+1}|\) between the spins. Two possible cases are for instance when the spins are either arranged near the maxima of the underlying oscillation of \(X^d(s,\bar{R})\) or on the other hand, when ferromagnetic and anti-ferromagnetic contributions are mostly canceling each other resulting in alternating signs of the interactions.

\subsection{A Pair of spatially separated spins}

\begin{figure}
\includegraphics[width=0.33\textwidth]{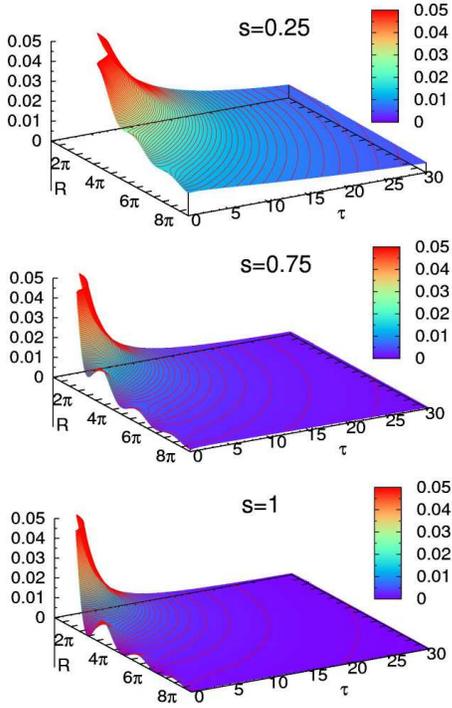}
\caption{\label{fig:kernel} Integral kernel \(K_{\beta}(\tau,\bar{R})\) at \(d=3\) for low imaginary time \(0 \leq \tau \leq 30\) and scaled distances \(0 \leq \bar{R} \leq 8\pi\) for different bath exponents (\(s=0.25\) (top), \(s=0.75\) (middle) and \(s=1\) (bottom))}
\end{figure}

In this section we study the ($N=2$)-MSBM with spatially separated spins
in a distance \(\bar{R}\). For a pair of spins, the action (\ref{eq:action}), separates in four parts \(S = S_1 + S_2 + 2S_{R}\), namely the self-energy for each spin
\begin{align}
\label{eq:action_S12}
S_{m} = \int_0^{\beta} \int_0^{\tau}  \sigma_{m}^z(\tau) \sigma_{m}^z(\tau') K_{\beta}(\tau - \tau', 0)  \mathrm{d}\tau' \mathrm{d}\tau
\end{align}
for $m=1,2$, and twice the interaction between the two spins
\begin{align}
\label{eq:action_SR}
S_{R} = \int_0^{\beta} \int_0^{\tau}  \sigma_{1}^z(\tau) \sigma_{2}^z(\tau') K_{\beta}(\tau - \tau', \bar{R})  \mathrm{d}\tau' \mathrm{d}\tau\;.
\end{align}
The Monte Carlo cluster algorithm (see Appendix~\ref{sec:algorithm}) can be adapted to the present case by taking all of these action into account. The key step is to evaluate the integral kernel \(K_{\beta}(\tau, R)\) (\ref{eq:kernel}) with the spectral function (\ref{eq:spectral_spasep}), which is modified by the spatial function \(f^d(\bar{R})\). The kernel must be non-negative in order to apply the algorithm, that is designed for ferromagnetic interaction only. The condition for that can be obtained by calculating the values of \(s\), for which \(K_{\beta}(\tau, R)\) begins to exhibits roots with respect to \(\tau\). At these onset values of \(s\), the roots are at \(\tau = 0\) and \(\tau=\beta\). In the limit \(\beta \rightarrow \infty\), the value of kernel at these points are
\begin{align}
K_{\beta}(0, R) = K_{\beta}(\beta, R) &= \int_{0}^{\omega_c} 2 \alpha \omega^{s-1} \omega_c^{1-s} f^d\left( \frac{\omega R}{v} \right) \frac{\beta \hbar }{2} \mathrm{d}\omega \nonumber \\
&= \frac{2 \omega_c^2 \alpha}{s+1} X^d\left(s+1, \bar{R}\right) \frac{\beta \hbar }{2}\;.
\end{align}
The function \(X^d(s,\bar{R})\) exhibits roots at \(s \geq s_{AF}^d\)  (Sec. \ref{sec:diss_ary_spa_sep}). This means, that the kernel is non-negative only for bath exponents \(s<s_{AF}^d-1\). If we restrict ourselves to the case of a three-dimensional bath, we are save to explore the complete (sub-)Ohmic regime (\(0< s \leq 1\)). The integral kernel is plotted in Fig.~\ref{fig:kernel} for several parameters for \(d=3\). At the \(s=1\), one can see the roots at \(\bar{R}=2\pi, 4\pi, 6\pi, \dots\).

\subsubsection{Phase transition point}
\begin{figure}
\includegraphics[width=0.45\textwidth]{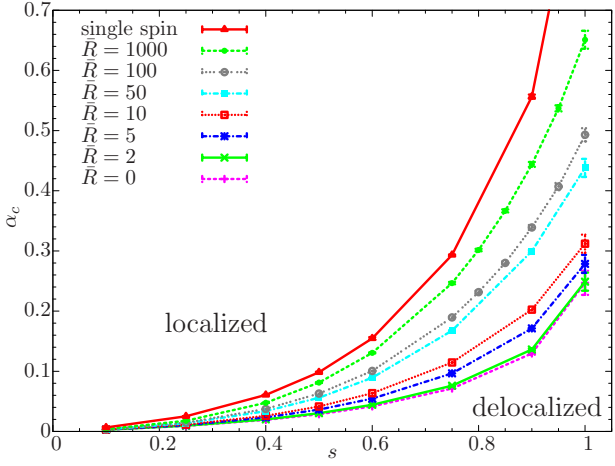}
\caption{\label{fig:phasediagram} 
Phase diagram of the $(N=2)$-MSBM with spatial separation for
various distances $\bar{R}$. The case \(\bar{R}=0\) corresponds to the MSBM with \(N=2\) from Sec. \ref{sec:nimp} and the case \(\bar{R}=\infty\) represents the common single spin-boson model.}
\end{figure}

As already known from \cite{McCutcheon2010, LeHur2010}, the transition point from delocalization to localization of two spins in a common bath is lowered in comparison to the single spin case. If one introduces a finite spatial separation, the transition point increases up to the single spin case, if \(\bar{R} = \infty\). We determined the phase diagram as in section \ref{secQMC} via finite-$\beta$-scaling of the fourth order cumulant \(Q\). Fig.~\ref{fig:phasediagram} shows the phase diagram for various distances. One can see, that the spins are influencing each other over relatively large distances. Even at \(\bar{R} = 1000\), the transition point is still clearly distinguishable from the single spins case. The localization is strongly enhanced by the presence of a second impurity in the bath. Note that the model neglects retardation effects, that are the more important, the further the spins are separated. Therefore, the bath mediated spin-spin influence at large distances may be less prominent in an real experimental setup.

\subsubsection{Order parameter and Spin-spin correlation}
\begin{figure}
\includegraphics[width=0.48\textwidth]{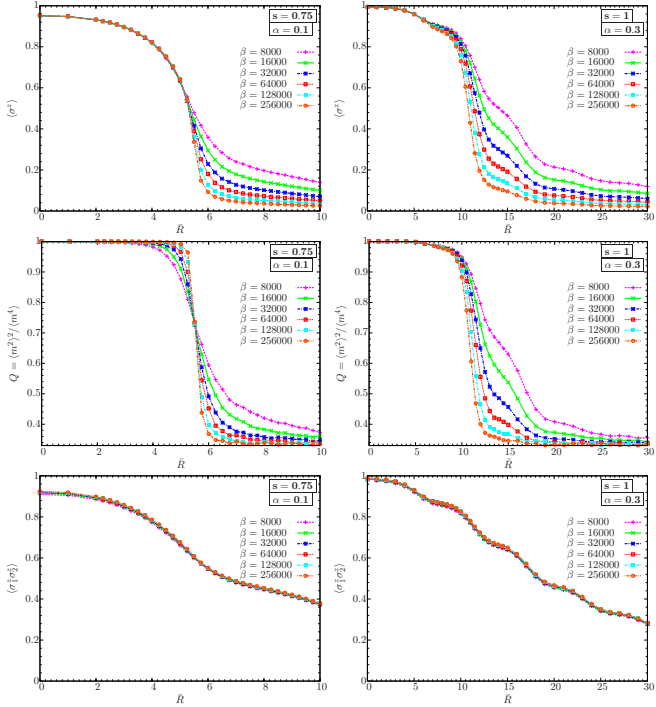}
\caption{\label{fig:beta_all} Order parameter \(\langle \sigma^z\rangle \) (top) fourth order cumulant \(Q\) (middle) and the correlation function \(\langle \sigma^z_1 \sigma^z_2 \rangle \) (bottom) versus the distance \(\bar{R}\) for different inverse temperatures. In the left column the parameter \(s=0.75\) and \(\alpha = 0.1\) is used whereas in the right column \(s=1\) and \(\alpha = 0.3\) is assumed.}
\end{figure}

We will now focus on the observables. First we will address extrapolation to zero-temperature. In Fig.~\ref{fig:beta_all} the order parameter \(\langle \sigma^z\rangle \), the fourth order cumulant \(Q\) and the correlation function \(\langle \sigma^z_1 \sigma^z_2 \rangle \) versus the distance \(\bar{R}\) is shown for several inverse temperature and for two different points in the parameter space, namely \(s=0.75, \alpha=0.1\) and \(s=1, \alpha=0.3\). There are two things to realize: The fourth order cumulants intersect for \(s=0.75\) and merge each other smoothly for \(s=1\), which is a characteristic, that the former case is ordinary continuous phase transition whereas the latter one is expected to be a phase transition of a Kosterlitz-Thouless type. The second things is, that the correlation function is nearly not affected by the finite size.

\begin{figure}
\includegraphics[width=0.48\textwidth]{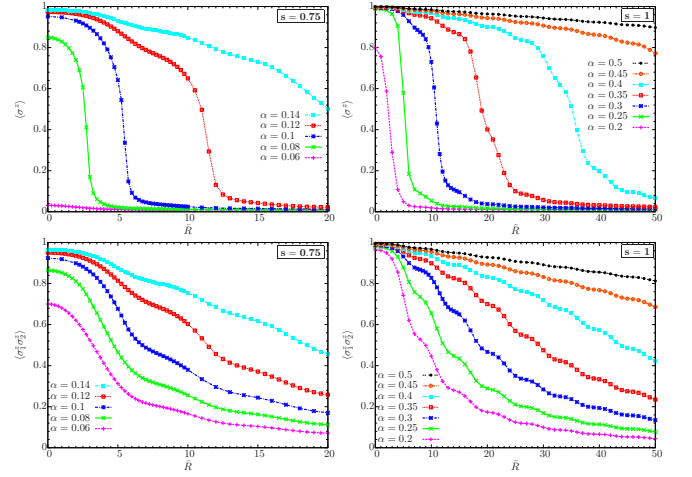}
\caption{\label{fig:alpha_all} Order parameter \(\langle \sigma^z\rangle \) (top) and the correlation function \(\langle \sigma^z_1 \sigma^z_2 \rangle \) (bottom) versus the distance \(\bar{R}\) for the parameter \(s=0.75\) (left) and the Ohmic spectrum \(s=1\) (right). The inverse temperature for all data is \(\beta = 256000\).}
\end{figure}

In Fig.~\ref{fig:alpha_all} we plot the order parameter and the correlation function versus the distance for several coupling strengths and for the two selected parameters \(s=0.75\) and \(s=1\). Basically the observables behave typically when driven through the phase transition by increasing the control parameter \(\bar{R}\). As a special feature, the curves show a underlying wiggling that is caused by the oscillations of the spatial dependent interaction (cf. Sec.~\ref{sec:diss_ary_spa_sep}). Also worth mentioning is the observation, that phase transition is not really recognizable by the correlation function curves despite the significant jump of the order parameter. This fact is however also the present without spatial separation as one can see Fig.~\ref{fig:obs_vs_alpha_sall}(b) for the same \(s\).

\section{Conclusions}
\label{sec:conclusion}

With the help of analytic considerations and large scale QMC 
simulations we have shown that the quantum phase transition in the 
(unbiased) multi-spin-boson model, describing $N$ independent 
two-level system  coupled to a common bosonic bath, falls in
the universality class of the single-spin-boson model,
i.e.\ is independent of the number of two-level systems
for any finite $N$. In the limit $N\to\infty$ the universality
class of the transition changes to mean-field behavior and
the system has a phase transition also at finite temperatures.
Consequently for large values of $N$ one observes a crossover
from mean-field to finite $N$ MSBM critical behavior for
decreasing temperature. 

The leading $N$-dependence of the critical bath coupling 
is shown to be $\alpha_c\simeq\frac{1}{N}\cdot\frac{s\Delta}{2\omega_c}
+\frac{a(s,\Delta)}{N^2}$, i.e.\ asymptotically the 
critical bath coupling is proportional to $1/N$. 
This confirms that the common bath mediates a ferromagnetic
interaction between the two-level systems, reduces fluctuations
and thus also the critical bath coupling. We showed
that the effective interaction between the two-level systems
can be quantified by the spin-spin correlation function,
which interpolates smoothly between mean-field behavior
for $s\to0$ (including a transition from uncorrelated 
to correlated two-level systems) and a simple ferromagnetically 
coupled two-spin system without bath for $s\gg1$. For
$s\gg1$ the short-ranged interactions in imaginary time 
still mediate an instantaneous ferromagnetic interaction
of strength $J=\alpha\omega_c/s$. For $s<1$ the behavior of
the connected spin-spin correlation function is reminiscent
of the entanglement entropy of the two-level systems: 
approximately linearly increasing with the bath coupling $\alpha$
in the delocalized phase $\alpha<\alpha_c$ and then rapidly 
decreasing in the localized phase ($\alpha>\alpha_c$).

For spatially separated two-level systems the critical bath coupling 
monotonously increases with the distance $R$ from the value for
the MSBM without spatial separation for $R=0$ to the value for the 
single-spin-boson model for infinite separation $R\to\infty$.
No indication is found for an alteration of the universality class of the transition by the
spatial separation. The order parameter as well as the spin-spin 
correlation function decreases systematically with the distance 
$R$ but shows superimposed oscillations.

The results reported in this paper are for the MSBM without a
direct ferromagnetic interaction between the two-level systems,
and one obvious question arises immediately: Will the universality
class of the transition change if a direct coupling between the 
two-level systems is introduced? NRG-results for a pair of spins ($N=2$)
reported in [\onlinecite{LeHur2010}] are confined to the transition 
triggered by the ferromagnetic coupling for fixed bath coupling
and predict exponents (e.g.\ $\beta=0.09$ for $s=0.9$ and
$\beta=0.2$ for $s=0.75$), which cannot be compared directly with 
the bath coupling triggered critical behavior of the MSBM 
(for which we found $\beta=0.037$ for $s=0.9$ and $\beta=0.23$ 
for $s=0.75$). An alternative view
was obtained recently\cite{Bonart2013} with a renormalization 
group calculation for a pair of spins each with its own bath,
which might behave differently than the case we are considering here.
Note that the quantum critical point of a chain of ferromagnetically coupled spins, each with its own bath, was studied in the limit $N\to\infty$ in Ref.~[\onlinecite{Werner2005}]. Again the question is, whether the 
reported universality class changes for a common bath. Certainly it will
be worthwhile to study these questions with the method that we used in 
this paper.

\appendix
\section{Derivation of the action}
\label{sec:deriv}
We are using the Caldeira-Leggett formulation \cite{CLM} for quantum dissipative systems, where the \(1/2\)-spins are represented by spinless particles at positions \(Q_m\) and momentum \(P_m\) in a symmetric double-well potential \(V(Q_m)\). The full Hamiltonian of \(N\) spatially separated spins in a common bosonic bath can written as
\begin{align}
H &= H_S + H_B + H_I + H_c
\end{align}
with
\begin{align}
H_S &= \sum_{m=1}^{N} \left(\frac{P_m^2}{2M} + V(Q_m) \right) \label{eq:H_S}\\
H_B &= \sum_{\kk=-\infty}^{\infty} \frac{{p_\kk}^2}{2m_\kk} + \sum_{\kk=-\infty}^{\infty} \frac{m_\kk \omega_\kk^2}{2}{x_\kk}^2\\
H_I &= \sum_{\kk=-\infty}^{\infty} c_\kk x_\kk \sum_{m=1}^N Q_m \exp(i \kk \!\cdot\! \rr_m)\\
H_C &= \sum_{\kk=-\infty}^{\infty} \frac{|c_\kk|^2}{2 m_\kk \omega_\kk^2} \sum_{m=1}^{N} \sum_{m'=1}^{N} Q_m Q_{m'} \exp(i \kk \!\cdot\! (\rr_m - \rr_{m'}))\;,
\end{align}
where \(H_S\) is the (inner) system part, \(H_B\) denoted the bath degrees of freedom, \(H_I\) is the interaction term and \(H_C\) the so-called counter-term, that compensates a global energy shift. The linear coupling between the particles and the bath mode is imposed by the coefficients \(c_{\kk}\). The common procedure is to expressing the partition function as a path integral \cite{Weissbook}
\begin{align}
Z \propto \int \prod_m \mathcal{D}[Q_m]\mathcal{D}[P_m] \prod_\kk \mathcal{D}[x_\kk]\mathcal{D}[p_\kk]  \exp \left( -S \right)\;.
\end{align}
and switching to Fourier space \(Q_m(\tau) = \sum_n Q_{m,n} \exp (i \omega_n \tau )\) (\(P_{m,n},p_{\kk,n}\) and \(x_{\kk,n}\) similarly), where \(\omega_n = 2 \pi n / \beta\) are  the Matsubara frequencies and \(\tau\) the imaginary time coordinate. The bath degrees of freedom and the particle momenta can now eliminated by Gaussian integration. The resulting interacting (non-local) part of the action is then given by
\begin{align}
S_I = \beta\sum_{m,m'=1}^N \left[\sum_{n=-\infty}^{\infty} \alpha(n,\rr_m \!-\! \rr_{m'})   Q_{m,n}Q_{m',-n} \right]
\end{align}
with the kernel
\begin{align}
\alpha(n,\mathbf{R}) &= \sum_{\kk} \frac{|c_\kk|^2 \omega_n^2 \exp( -i \kk \!\cdot\! \mathbf{R})}{2 m_\kk \omega_\kk^2 (\omega_n^2 + \omega_\kk^2)}
\end{align}
Transforming back by means of the identity \cite{Gradshteyn}
\begin{align}
\sum_n \frac{\omega_k^2}{(\omega_n)^2 + \omega_k^2} e^{i\omega_n (\tau - \tau')} = \frac{\beta \hbar}{2} \omega_k \frac{\cosh\left(\frac{\beta \omega_k}{2}\left(1- \frac{2|\tau-\tau'|}{\beta}\right) \right)}{\sinh\left( \frac{\beta}{2} \omega_k\right)}
\end{align}
leads to the representation of Eqs.~(\ref{eq:action}),~(\ref{eq:kernel}),~(\ref{eq:spectral_spasep_sum}) with \(|\lambda_\kk|^2 = |c_\kk|^2/(2 m_\kk \omega_\kk)\). The (local) system part (\ref{eq:H_S}) becomes in the spin-boson limit  a two-state system whose path integral representation is a Poissonian distribution of an even number of kinks \cite{FarhiGutmann,RiegerKawashima1999,Krzakala}.

\section{The continuous time cluster QMC algorithm}
\label{sec:algorithm}

\begin{figure}
\includegraphics[width=0.45\textwidth]{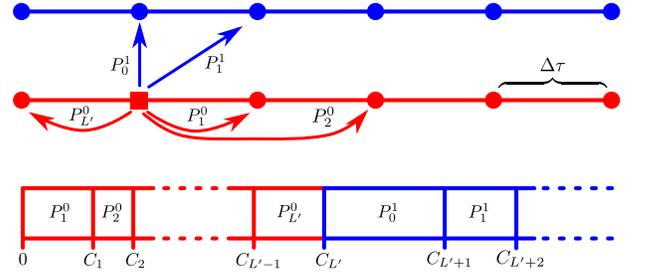}
\caption{\label{fig:paperalgo}Sketch the arrangement of the skipping probabilities \(P_k^n\) and the array of the cumulative probabilities \(C_k\) for two worldlines. The neighbors are counted to the right meaning the the left neighbor of the same worldline has the label \(L'\). In this example, the reference spin to which the bonds shall connected to is depicted by the square. The skipping probabilities to the spins of the second worldline are concatenated the cumulative probability array. The draw of a random number \(C_{L'+1} < g \le C_{L'+2}\) would correspond to the selection of the nearest neighbor spin to the right of the \emph{other} worldline.}
\end{figure}

In this appendix we describe the essential steps of the QMC procedure we are using. To demonstrate the principle of the algorithm we start with a discretized (Trotter) representation of the action
\begin{align}
\label{eq:algo_discret_S}
S=-\frac{1}{2} \sum_{i,j=1}^{L'} J_{|i-j|} s_i s_j - K \sum_{i=1}^{L'} s_i s_{i+1}
\end{align}
with a ferromagnetic long-range
and a nearest neighbor part
\begin{align}
\label{eq:algo_JuK}
J_{i} = \frac{\Delta \tau^2}{2 \beta} K_{\beta}\left(i \Delta \tau\right)  \quad \mbox{and} \quad K=-\frac{1}{2} \log \left(\tanh\left(\frac{\Delta}{2} \cdot \Delta \tau\right)\right)
\end{align}
and perform the continuous time limit afterwards. Within this discretization, the worldline is divided into spins with infinitesimal lattice spacing \(\Delta \tau=\beta/L' \rightarrow 0\) in contrast to the approximative discretization of Sec.~\ref{sec:QPT_MSBM_theory} which would correspond to \(\Delta \tau=2/\Delta\). Cluster Monte Carlo methods for classical Ising spins models involve a cluster building procedure, that adds equally aligned (up or down) spins \(s_i\) and \(s_j\) with a bond activation probability (BAP)
\begin{align}
\label{eq:algo_BAP}
p_{|i-j|} =  1 - \exp \left(-2 J_{|i-j]} - 2 K \delta_{1,|i-j|}\right)
\end{align}
to a cluster \cite{SwendsenWang1987}.
The two parts of the interaction can be treated separately in the cluster building process since the separation of long-range and nearest-neighbor BAPs
\begin{align}
p_{i}^\mathrm{(lr)} &=  1-\exp \left(-2 J_{i} \right) \\
p_{i}^\mathrm{(nn)} &=  1-\exp \left(-2 K \delta_{1,i}\right)
\end{align}
yield the correct global BAP (\ref{eq:algo_BAP}) via
\begin{align}
p_{i} &= p_{i}^\mathrm{(lr)} + p_{i}^\mathrm{(nn)} - p_{i}^\mathrm{(lr)} p_{i}^\mathrm{(nn)}\\
&= 1 - \left(1 - p_{i}^\mathrm{(lr)}\right) \left(1 - p_{i}^\mathrm{(nn)} \right)\;.
\end{align}
We will now focus on the long-range interaction and will describe, how the cluster building process can performed quickly \cite{Luijten1997}. The probability for activating a bond to the \(k-\)th neighbor without making any bond to neighbors in between is
\begin{align}
\label{eq:algo_P}
P_k = p_k^\mathrm{(lr)} \prod_{i=1}^{k-1}\left(1-p_{i}^\mathrm{(lr)}\right)
\end{align}
and will be called skipping probability in the following.
Note, that only the equal alignment of the reference spin and its \(k-\)th neighbor is required to add the candidate \(s_k\) to the cluster, so that the spins \(s_1, \dots s_{k-1}\) can be skipped without being touched. The cumulative probability
\begin{align}
C_{l} = \sum_{k=1}^{l} P_k = 1 - \exp \left(-2 \sum_{i=1}^{l} J_{i} \right)
\end{align}
enables one to draw directly (without rejections) the next candidate \(l\) for making a bond to by a formal inversion
\begin{align}
l=C^{-1} \left(g\right)\;,
\end{align}
where \(g=\mathrm{ran}(0,C(L'))\) is a random number uniformly 
distributed between $0$ and $C(L')$.
This rejection-free procedure can build up for arbitrary ordered probabilities. If we label the different worldlines in a way, that we can address them by a fixed and order sequence, every worldline possesses its well-defined neighbor worldlines. The extended skipping probability for activating a bond to the \(l-\)th neighbor spin but of the \(n-\)th neighbor worldline where all spins of all worldlines in between have been skipped can defined by (cf. Fig.~\ref{fig:paperalgo})
\begin{align}
\label{eq:algo_P2}
P_k^n = \left( \prod_{m=0}^{n-1} \prod_{i=1}^{L'} \left(1-p_{i}^\mathrm{(lr),m}\right)  \right)p_k^\mathrm{(lr),n} \prod_{i=1}^{k-1}\left(1-p_{i}^\mathrm{(lr),n}\right)\;.
\end{align}

Note, that for \(n=0\) the first product is empty and (\ref{eq:algo_P}) is recovered.
We can now append these probabilities (\ref{eq:algo_P2}) to the cumulative probability 
\begin{align}
C_{n\cdot L' + l} = \sum_{m=0}^n\sum_{k=1}^{l} P_k^m
\end{align}
and process as in the single worldline case.
For later purposes a cumulative probability for groups of spin has to be defined. Considering a certain number \(r\) with the labels \(s_{-r+1}, \dots s_0\) of adjacent and equally orientated spins. The probability for activating at least one bond between these spins and another one \(s_i\) (\(i>0\)) is
\begin{align}
p_{i}^\mathrm{(lr)}(r) &= 1 - \prod_{j=0}^{r-1}\left(1 - p_{i+j}^\mathrm{(lr)}\right) \nonumber \\
&= 1-\exp \left(-2 \sum_{j=0}^{r-1} J_{i+j} \right)\;.
\end{align}
Based on this probability modified skipping probabilities and cumulative probabilities can be defined in the same way as described above leading to
\begin{align}
\label{eq:algo_C2}
C_{l,r} = 1 - \exp \left(-2 \sum_{i=1}^{l} \sum_{j=0}^{r-1} J_{i+j} \right)\;.
\end{align}

We will now perform the continuous time limit \({\Delta \tau \rightarrow 0}\), which means, that a sequence of equally aligned spins \(s_i,s_{i+1} \dots, s_j\) is represented as a continuous segment \([s(i),s(j)]\) of length \(t = |j-i|\).
The nearest-neighbor BAP becomes \(p_{i}^\mathrm{nn} = 1 - \Delta/2\cdot \Delta \tau\) in the \(\Delta \tau \rightarrow 0\) limit and the probability for adding \emph{all} equally aligned spins up to a distance \(t\) is
\begin{align}
\label{eq:algo_p_cont}
p^\mathrm{\mathrm{seg}}(t) = \lim_{\Delta \tau \rightarrow 0}\left(1 - \frac{\Delta}{2} \Delta \tau  \right)^{t/\Delta \tau} = \exp\left(-\frac{\Delta}{2} t\right)\;.
\end{align}
The cumulative long-range probability can expressed by inserting (\ref{eq:algo_JuK}) in (\ref{eq:algo_C2}) and reads
\begin{align}
\label{eq:algo_C3}
C(l,r) &= 1 - \exp \left(-\frac{1}{\beta}\int_{0}^{l} \mathrm{d}\tau \int_{-r}^{0}  \mathrm{d}\tau' K_{\beta}(\tau - \tau') \right)\;.
\end{align}
This is the cumulative distribution probability for activating a bond between a reference segment at \([-r,0]\) and the segment which is located at a distance \(l\). Note, that for multiple worldlines  \(K_{\beta}(\tau-\tau')\) has the meaning of a concatenated function of the different \(K_{\beta}\)'s. Without giving the derivation, the cumulative distribution function for activating the segment at the distance \(l'\) under the condition, that the point \(l\) has been selected before is
\begin{align}
\label{eq:algo_C4}
C_l(l',r) &= 1 - \exp \left(-\frac{1}{\beta}\int_{l}^{l'} \mathrm{d}\tau \int_{-r}^{0}  \mathrm{d}\tau' K_{\beta}(\tau - \tau') \right)\;.
\end{align}
For the required inversion we recast (\ref{eq:algo_C4}) to the form
\begin{align}
\label{eq:algo_T-T}
- \log \left( 1 - C(l,r) \right) = T_r(l') - T_r(l)
\end{align}
where the function \(T_r(l) = (K_{\beta}^{(2)}(l+r) - K_{\beta}^{(2)}(l))/\beta\) contains the second integral of the kernel \(K_{\beta}(t)\).

\begin{figure}
\includegraphics[width=0.45\textwidth]{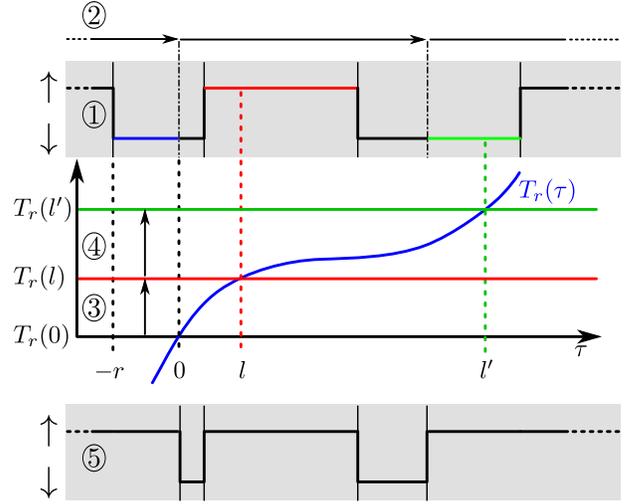}
\caption{\label{fig:paperalgo_mcstep} A exemplary part with four kinks of an arbitrary worldline is shown in \((1)\). In step \((2)\) exponential distributed random numbers according to (\ref{eq:algo_p_cont}) are drawn and sequentially inserted to the worldline (two dash-dotted lines). These ''cuts'' together with the already existing kinks divide the worldline into segments. Every segment is able to initiate a cluster building process. Without loss of generality the first (\(\downarrow\))-segment to the left (blue) is chosen and its endpoint is shall define the origin of coordinates. Drawing an exponential distributed random number  in \((3)\) gives the value \({T_{r}(l) = T_{r}(0) - \log \left( \mathrm{ran}(0,1) \right)}\) (c.f (\ref{eq:algo_T-T})). Determine numerically the point \(l\) and its corresponding segment (red), leads to the next candidate for adding to the cluster, which will be rejected in the present case due to its incompatible (\(\uparrow\))-orientation. Proceeding likewise with \(l\) as the starting point leads to another candidate (green) in step \((4)\), which is this time actually added to the cluster. Finally, if this cluster is flipped and the redundant ''cuts'' are removed the worldline would looks like in \((5)\).}
\end{figure}

In practice, we pre-calculate \(K^{(2)}_{\beta}(\tau)\) by means of an adaptive spline interpolation for quicker evaluation during the simulation. The essential steps of the algorithm are summarized in Fig.~\ref{fig:paperalgo_mcstep}.

\section{Low temperature mean-field solution}
\label{sec:mean-field_solution}

We start with the discretized form of the partition function (\ref{eq:partition_discret}) with the abbreviation for the interaction (\ref{eq:partition_binom_defs})
\begin{align*}
Z = \sum_{\{s_{m,i}\}} \exp \left[ \frac{1}{2 N} \sum_{m,m'}^N \sum_{i,j}^L  J_{i,j}  s_{m,i} s_{m',j}\right]\;.
\end{align*}
Introducing continuous variables \(m_i\) by means of delta functions leads to
\begin{align*}
Z &= \sum_{\{s_{m,i}\}} \exp \left[ \frac{N}{2}  \sum_{i,j}^L J_{i,j} \left(\sum_{m=1}^N \frac{s_{m,i}}{N} \right) \left(\sum_{m'=1}^N \frac{s_{m',j}}{N} \right)  \right]\\
  &= \sum_{\{s_{m,i}\}} \int \prod_{i=1}^L \left[\mathrm{d}m_i \delta \left(m_i - \sum_{m=1}^N \frac{s_{m,i}}{N} \right)\right] \exp \left[ \frac{N}{2}  \sum_{i,j}^L J_{i,j} m_i m_j \right]\;.
\end{align*}
The delta functions can substituted by
\begin{align*}
\delta \left(m_i - \sum_{m=1}^N \frac{s_{m,i}}{N} \right) &= N \delta \left(N m_i - \sum_{m=1}^N s_{m,i} \right) \\
 &= \frac{N}{2 \pi} \int \exp \left( i \hat{m}_i \left(N m_i - \sum_{m=1}^N s_{m,i} \right)\right) \mathrm{d}\hat{m}_i\;,
\end{align*}
which leads to
\begin{align*}
Z &= \int \prod_{i=1}^L \left[\frac{\mathrm{d}\hat{m}_i \mathrm{d}m_i }{2 \pi N^{-1}}\right]  \sum_{\{s_{m,i}\}} \exp \left( i N \sum_{i=1}^L \hat{m}_i  m_i \right) \times \\
&\phantom{=} \qquad \exp \left(-i \sum_{i,m} \hat{m}_i s_{m,i}\right)  \exp \left[ \frac{N}{2}  \sum_{i,j}^L J_{i,j} m_i m_j \right]
\end{align*}
Now, the trace over the spin variables \(s_{m,i}\) can performed directly
\begin{align*}
Z &= \int \prod_{i=1}^L \left[\frac{\mathrm{d}\hat{m}_i \mathrm{d}m_i }{2 \pi N^{-1}}\right]  \prod_{i} \left[2 \cos \left(\hat{m}_i \right) \right]^N \times \\
&\phantom{=} \qquad \exp \left( i N \sum_{i=1}^L \hat{m}_i  m_i \right) \exp \left[ \frac{N}{2}  \sum_{i,j}^L J_{i,j} m_i m_j \right]\\
&= \int \prod_{i=1}^L \left[\frac{\mathrm{d}\hat{m}_i \mathrm{d}m_i }{2 \pi N^{-1}}\right]  \exp \left[ N \cdot f (\{m_i\},\{\hat{m}_i\}) \right]\;,
\end{align*}
where in the last line the function
\begin{align*}
 f (\{m_i\},\{\hat{m}_i\}) = \ln (2 \cos \left(\hat{m}_i \right)) + i \sum_{i=1}^L \hat{m}_i  m_i  +  \frac{1}{2}\sum_{i,j}^L J_{i,j} m_i m_j
\end{align*}
was defined. In the limit \(N \rightarrow \infty\), a saddle point integration leads finally to the common mean-field relation for the magnetization
\begin{align}
\label{eq:mean_field_solution}
\frac{\partial f}{\partial m_i} = 0 = \frac{\partial f}{\partial \hat{m}_i} \; \Rightarrow \; m_i = \tanh \left(\sum_{j=1}^L J_{i,j} m_j \right)\;.
\end{align}
In equilibrium, all \(m_i=m\) are equal and the phase transition point can calculated from (\ref{eq:mean_field_solution}) by the condition \(\sum_j J_{i,j} = 1\). Substituting back \(J_{i,j}\) (\ref{eq:partition_binom_defs}), one obtains
\begin{align}
\label{eq:mean_field_solution2}
\sum_{j=1}^{L} J_{i,j} = \int_0^{\beta}  \frac{1}{\Delta \beta} \tilde{K}_{\beta}\left(\tau- \tau'\right)   \mathrm{d}\tau = \frac{2 \omega_c \tilde{\alpha}}{\Delta s}
\end{align}
where the replacement of the sum by the integral is valid for \(\mathcal{O}(1) \ll \Delta \beta\). Setting Eq.~(\ref{eq:mean_field_solution2}) equal to \(1\) recovers the critical point.



\begin{thebibliography}{}

\bibitem{Weissbook}
U. Weiss, Quantum Dissipative Systems, 4th ed. (World Scientific, Singapore, 2012).

\bibitem{CLM}
A. J. Leggett, S. Chakravarty, A. T. Dorsey, M. Fisher, A. Garg, and W. Zwerger, Rev. Mod. Phys. \textbf{59}, 1 (1987).

\bibitem{Winter2009}
A. Winter, H. Rieger, M. Vojta, and R. Bulla, Phys. Rev. Lett. \textbf{102}, 030601 (2009).

\bibitem{Alvermann2009}
A. Alvermann and H. Fehske, Phys. Rev. Lett. \textbf{102}, 150601 (2009).

\bibitem{Zhang2010}
Y. Zhang, Q. Chen, and K. Wang, Phys. Rev. B \textbf{81}, 121105 (2010).

\bibitem{Stamp1998}
M. Dub\'{e} and P. Stamp, Int. J. Mod. Phys. B \textbf{12}, 1191 (1998).

\bibitem{Governale2001}
M., Governale, M. Grifoni, and G. Sch\"on, Chem. Phys. \textbf{268}, 273 (2001)

\bibitem{Thorwart2002}
M. Thorwart and P. H\"anggi, Phys. Rev. A \textbf{65}, 012309 (2002)

\bibitem{Garst2004}
M. Garst, S. Kehrein, T. Pruschke, A. Rosch, and M. Vojta, 
Phys. Rev. B \textbf{69}, 214413 (2004).

\bibitem{Nagele2008}
N\"agele, G. Campagnano, and U. Weiss, New J. Phys. \textbf{10}, 115010 (2008).

\bibitem{Nagele2010}
N\"agele and U. Weiss, Physica E Amsterdam \textbf{42}, 622 (2010).

\bibitem{LeHur2010}
Peter P. Orth, David Roosen, Walter Hofstetter, and Karyn Le Hur, Phys. Rev. B \textbf{82}, 144423 (2010).

\bibitem{McCutcheon2010}
Dara P. S. McCutcheon, Ahsan Nazir, Sougato Bose, and Andrew J. Fisher, Phys. Rev. B \textbf{81}, 235321 (2010).

\bibitem{Bonart2013}
J. Bonart, Phys. Rev. B {\bf 88}, 125139 (2013).

\bibitem{Werner2005}
P. Werner, K. V\"{o}lker, M. Troyer, and S. Chakravarty, Phys. Rev. Lett. \textbf{94}, 047201 (2005).

\bibitem{Cugliandolo2005}
L. F. Cugliandolo, G.S. Lozano, and H. Lozza,
Phys. Rev. B \textbf{71}, 224421 (2005).

\bibitem{Schehr2006}
G. Schehr and H. Rieger, Phys. Rev. Lett. \textbf{96}, 227201 (2006). 
G. Schehr and H. Rieger, J. Stat. Mech., 04012 (2008).

\bibitem{Vojta2008}
J. A. Hoyos and T. Vojta,
Phys. Rev. Lett. {\bf 100}, 240601 (2008),
Phys. Rev. B {\bf 85}, 174403 (2012).

\bibitem{Baxterbook}
Rodney J. Baxter, Exactly solved models in statistical mechanics, (Academic Press, London and New York, 1982).

\bibitem{HouTong2010}
Y.-H. Hou and N.-H. Tong, Eur. Phys. J. B \textbf{78}, 127 (2010).

\bibitem{Fisher1972}
Michael E. Fisher, Shang-Keng Ma, and B. G. Nickel,
Phys. Rev. Lett. {\bf 92}, 917 (1972).

\bibitem{Luijten1997}
Erik Luijten and Henk W. J. Bl\"{o}te, Phys. Rev. B \textbf{56}, 8945 (1997).

\bibitem{fn:seg_length}
In an actual simulation, the real mean length of a segment will be larger then \(2/\Delta\), if the long-range action is present. But since the influence of this long-range action will be incorporated separately, one has to consider the bare characteristic segment length for discretization.

\bibitem{Luijten1996}
E. Luijten, H. W. J. Bl\"{o}te, and K. Binder, Phys. Rev. E \textbf{54}, 4626 (1996).

\bibitem{Aharony}
A. Aharony, Phase Transitions and Critical Phenomena, edited by C. Domb and M. S. Green (Academic
Press, London, 1976), vol. 6, chap. 6.

\bibitem{Scalapino1981}
J. Bhattacharjee, S. Chakravarty, J. L. Richardson, and D. J. Scalapino, Phys. Rev. B \textbf{24}, 3862 (1981).

\bibitem{Messingfeld}
E. Luijten and H. Me{\ss}ingfeld, Phys. Rev. Lett. \textbf{86}, 5305 (2001).

\bibitem{Carrbook}
Karyn Le Hur, Quantum Phase Transitions in Spin-Boson Systems. In Lincoln D. Carr (Ed.), Understanding Quantum Phase Transitions (Rev. ed., pp. 217-240). Boca Raton, CRC Press, 2011.

\bibitem{Mahan}
Mahan, G. D., Many-Particle Physics, 3nd ed. (Plenum, New York, 2000).

\bibitem{Zell2009}
T. Zell, F. Queisser, and R. Klesse, Phys. Rev. Lett. \textbf{102}, 160501 (2009).

\bibitem{Gradshteyn}
I. S. Gradshteyn and I. M. Ryzhik, Table of Integrals, Series, and Products. Edited by A. Jeffrey and D. Zwillinger. Academic Press, New York, 7th edition, 2007

\bibitem{SwendsenWang1987}
R. H. Swendsen and J. S. Wang, Phys. Rev. Lett. \textbf{58}, 86 (1987).

\bibitem{FarhiGutmann}
E. Farhi, S. Gutmann, Ann. Phys., \textbf{213} (1992), 182 (1992).

\bibitem{RiegerKawashima1999}
H. Rieger and N. Kawashima, Eur. Phys. J. B \textbf{9}, 233 (1999).

\bibitem{Krzakala}
F. Krzakala, A. Rosso, G. Semerjian, and F. Zamponi, Phys. Rev. B \textbf{78}, 134428 (2008).

%
%
%

\end{thebibliography}
\end{document}